\documentclass[12pt]{iopart}

\usepackage{iopams}
\usepackage{graphicx}
\usepackage{hyperref}
\hypersetup{
    colorlinks=true,
    linkcolor=blue,
    filecolor=magenta,      
    urlcolor=blue,
    citecolor=black,
}
\begin{document}

\title[Robust model benchmarking and bias-imbalance in data-driven materials science]{Robust model benchmarking and bias-imbalance in data-driven materials science: a case study on MODNet}
\author{Pierre-Paul De Breuck, Matthew L. Evans and Gian-Marco Rignanese}

\address{Université catholique de Louvain (UCLouvain), Institute of Condensed Matter and Nanosciences (IMCN), Chemin des Étoiles 8, B-1348 Louvain-la-Neuve, Belgium}
\ead{gian-marco.rignanese@uclouvain.be}
\vspace{10pt}
\begin{indented}
\item[]\today
\end{indented}

\begin{abstract}
As the number of novel data-driven approaches to material science continues to grow, it is crucial to perform consistent quality, reliability and applicability assessments of model performance.
In this paper, we benchmark the Materials Optimal Descriptor Network (MODNet) method and architecture against the recently released MatBench v0.1, a curated test suite of materials datasets.
MODNet is shown to outperform current leaders on 6 of the 13 tasks, whilst closely matching the current leaders on a further 2 tasks; MODNet performs particularly well when the number of samples is below 10,000.
Attention is paid to two topics of concern when benchmarking models. 
First, we encourage the reporting of a more diverse set of metrics as it leads to a more comprehensive and holistic comparison of model performance.
Second, an equally important task is the uncertainty assessment of a model towards a target domain.
Significant variations in validation errors can be observed, depending on the imbalance and bias in the training set (i.e., similarity between training and application space). By using an ensemble MODNet model, confidence intervals can be built and the uncertainty on individual predictions can be quantified.
Imbalance and bias issues are often overlooked, and yet are important for successful real-world applications of machine learning in materials science and condensed matter.
\end{abstract}

%
%
 \submitto{\JPCM}
%
%
%
\section{Introduction}

Functional materials are at the heart of many technological advances and are therefore essential to the growth of next-generation innovative and sustainable technologies~\cite{mageeQuantificationRoleMaterials2012}.
The capability to predict structure- or composition-property relationships would further accelerate development and when coupled with the proliferation of reliable materials simulations~\cite{lejaeghere2016reproducibility}, open datasets~\cite{himanen2019}, and high-throughput experimentation, machine learning (ML) has emerged as a particularly powerful approach~\cite{butler2018machine,Schmidt2019}.
Complex properties can indeed be predicted by surrogate models in a fraction of time with close to quantum accuracy, allowing for a much faster screening of materials; multi-fidelity models can go further and render these predictions more directly comparable to experimental measurements \cite{chen2021learning}.
For instance, Oliynyk~\textit{et al.} proposed a model that identifies Heusler compounds~\cite{oliynykHighThroughputMachineLearningDrivenSynthesis2016}.
From their study, novel predicted candidates from the MRu$_2$Ga and RuM$_2$Ga (M=Ti-Co) composition space were synthesised and confirmed to be Heusler compounds.
Another example concerns the work of Stanev~\textit{et al.}, which aimed to discover high-$T_c$ superconductors, by constructing an estimator for critical temperature $T_c$ based on the composition alone~\cite{stanevMachineLearningModeling2018}.
Databases of reported synthesised compositions were screened for new superconductors, yielding 35 non-cuprate and non-ferrous oxides as candidate compositions.
These examples, just two among many~\cite{butler2018machine,Schmidt2019}, show the impact of ML techniques in materials science and condensed matter, and as such it is considered as the fourth paradigm of materials science~\cite{agrawalPerspectiveMaterialsInformatics2016}.

Many ML approaches are being actively developed to accurately predict compound-property relationships.
They differ in the amount, type and richness of descriptors and model complexity, as well as adopting other learning paradigms such as transfer or multi-task learning.
As a result, they exhibit very different characteristics, and no universally superior algorithm exists~\cite{wolpertNoFreeLunch1997,wolpertCoevolutionaryFreeLunches2006}.
Firstly, the input type and thus applicability can be very different; some models are architecturally restricted to specific descriptors, e.g., graph models require atomic coordinates to make predictions~\cite{xieCrystalGraphConvolutional2018,chenGraphNetworksUniversal2019}.
Secondly, the ability for models to generalise (i.e., the error on unseen samples) strongly depends on the amount of available data, with different approaches succeeding in contrasting data-poor or data-rich regimes.
Often, a model or architecture that performs very well on a large dataset can be disappointing when applied on a small dataset (and \emph{vice versa})~\cite{debreuckMaterialsPropertyPrediction2021,dunnBenchmarkingMaterialsProperty2020}.

Given the number and variety of approaches being adopted, it is important that the field follows a standard benchmarking pipeline, inspired by other fields (e.g., ImageNet for computer vision~\cite{dengImageNetLargescaleHierarchical2009}).
Too often, models are tested on only a small number of datasets, with unclear or unrepeatable validation and testing procedures.
Significant variation in generalisation performances can be observed for different hold-out sets, as will be shown.
To draw a fair comparison between the two competing models, they should not just be benchmarked on the same dataset, but also cross-validated on the same training and testing subsets (typically, a 5-fold cross-validation with fixed random seed).
Similarly, hyperparameter choices are often opaque and should instead follow a reproducible pipeline.
Every model has advantages and disadvantages and testing a diverse range of tasks shows that no model is universally superior, but rather performance of a given architecture is a function of the amount of data, feature availability, the complexity of the target space and the computational resources available for tuning.
These reasons motivate the creation of a standard benchmark. Dunn~\textit{et al.} recently released the MatBench v0.1 test suite~\cite{dunnBenchmarkingMaterialsProperty2020} which consists of 13 materials science-specific prediction tasks, ranging from small ($\sim$300) to large ($\sim$150,000) datasets, both experimental and computational, with target properties covering a wide range of materials phenomena (mechanical, electronic, and thermodynamic), with both regression and classification tasks.

Convinced by this standardised approach, we apply the MatBench v0.1 benchmark to the recently reported Material Optimal Descriptor Network (MODNet) \cite{debreuckMaterialsPropertyPrediction2021} developed by some of the present authors. 
Overall, the performance of MODNet is quite competitive; out of the 13 tasks, MODNet is able to outperform (by more than 2\%) the current leader on 6 tasks or closely match it (within $\pm$2\%) on a further 2 tasks.
Of the remaining 5 tasks, the 4 largest datasets were not attempted as it has been shown that graph networks will dominate in this regime~\cite{debreuckMaterialsPropertyPrediction2021}. 
For the phonon DOS task, MODNet performs around 5\% worse than current leader when trained using automated hyperparameter optimisation, but this can be improved (and reversed) if the models are hand-tuned \emph{post hoc}.
As such, MODNet provides a promising alternative on small to medium-sized datasets to the automated but complex pipelines trained with Automatminer \cite{dunnBenchmarkingMaterialsProperty2020} and the data-intensive graph convolutional neural networks CGCNN \cite{xieCrystalGraphConvolutional2018} and MEGNet \cite{chenGraphNetworksUniversal2019}.

Furthermore, our testing reveals a variance in the generalisation error between cross-validation folds.
In the second part of this paper, our aim is to predict this (sometimes abrupt) change in prediction error by using an ensemble of MODNet models, with a discussion of the importance of imbalance and bias in the dataset, and how model performance can be appropriately reported within these different regimes.
This paves the way to a better understanding of a model's applicability for a particular target domain, allowing the estimation of the prediction error of unseen samples.

\section{Methods}


The Material Optimal Descriptor Network (MODNet) is an open source framework for predicting materials properties from primitives such as composition or structure~\cite{debreuckMaterialsPropertyPrediction2021}.
It consists of a feedforward neural network fed with a limited number of descriptors, derived from chemical, physical, and geometrical considerations (see~\ref{appendix:MODNet}).
MODNet was designed to make the most efficient use of data for tasks where large, internally consistent training sets are prohibitively difficult or expensive to obtain; we observe that most existing material datasets are limited in this way, particularly those derived from experimental results.
In order to have the best possible performance at low size, three key aspects are needed.

First, our observation showed that chemical and geometrical descriptors can provide more information than a raw graph representation. To this end, MODNet makes use of many of the existing descriptors designed in the community as implemented in the Matminer package~\cite{WardMatminer}.
By giving the model descriptors derived from chemical, physical, and geometrical considerations, part of the learning is already done as they exploit existing chemical knowledge.
In contrast, more complex models such as graph networks or embedding methods are often initialised with only atomic numbers and bond lengths~\cite{chenGraphNetworksUniversal2019, xieCrystalGraphConvolutional2018}.
Graph models are able to bootstrap this chemical knowledge \textit{ex nihilo}, and perform highly accurate property predictions but only when a large ($\sim10^4$) and sufficiently diverse dataset is available.

Second, the smaller the dataset, the greater the importance of feature selection for tackling the curse of dimensionality~\cite{jovicReviewFeatureSelection2015}.
For instance, a previous application of MODNet to vibrational thermodynamics~\cite{debreuckMaterialsPropertyPrediction2021} showed that an average 12\% improvement in error can be obtained by applying careful feature selection.
For feature selection, MODNet employs an iterative procedure based on a relevance-redundancy criterion (measured through the normalised mutual information between all pairs of features, and between features and targets), see~\ref{appendix:MODNet}.

Third, combined models that integrate other information in the form of transfer learning, multi-task/joint learning, multi-fidelity methods \cite{kauweExtractingKnowledgeDFT2020, chen2021learning}, or any other form, can mitigate the lack of data for a given target.
Indeed, when a model combines different properties for the same compound, a more general representation can be formed in the hidden layers that more efficiently uses the information provided to the learning algorithm and increasing model performance for a given dataset size.
With MODNet, one can learn on multiple properties by using a tree-like architecture, see Figure~\ref{fig:model_schematic} in~\ref{appendix:MODNet}.
The similarity between target properties can be used to decide where the tree splits, i.e., the layer up to which properties share an internal representation. 
This approach is particularly effective when learning parametric data that exhibit, for example, temperature or pressure dependence.

MODNet has been trained and tested on both single and multi-property tasks with excellent performance compared to contemporary methods for datasets containing 10,000 samples or fewer.
For instance, on the refractive index dataset of Naccarato~\textit{et al.}~\cite{naccaratoSearchingMaterialsHigh2019} (4040 samples), MODNet achieves a mean absolute error of 0.051 (n.b., this dataset is distinct from MatBench refractive index task).
A multi-target MODNet was also developed to study temperature-dependent results of the vibrational thermodynamics dataset of Petretto~\textit{et al.}~\cite{petrettoHighthroughputDensityfunctionalPerturbation2018} (1265 samples). 
A single model was simultaneously trained on the vibrational entropy, enthalpy, specific heat and Helmholtz energy across a wide temperature range. 
This model achieved an average error of 0.009\,meV/K/atom on the room temperature vibrational entropy ($S^\circ$), four times lower than previous studies, with joint learning contributing to an error reduction of 8\% compared to the best MODNet model trained on $S^\circ$ alone~\cite{debreuckMaterialsPropertyPrediction2021}.

Since the initial release (v0.1.5) of MODNet as reported in~\cite{debreuckMaterialsPropertyPrediction2021}, several improvements to the software framework have been implemented in the version used in this paper (v0.1.9).

First, an ensemble MODNet model has been implemented and used for this work. It consists of a set of MODNet models, all trained on the same task, but with independently optimised hyperparameters, training set and initial weights. In particular, we choose to take the 5 best architectures for each inner training fold. Moreover, multiple copies of each individual architecture are trained via bootstrap resampling of the given training set fold, resulting in a 125-model ensemble. It is based on the \emph{Deep Ensemble} framework~\cite{lakshminarayananSimpleScalablePredictive2017}, and the bootstrapping technique, that has been showed to have consistent advantage in out-of-domain epistemic uncertainty calibration and generalization error when applied on the chemical space~\cite{scaliaEvaluatingScalableUncertainty2020}.
This final model has the advantage to predict a distribution of values, from which statistics such as the mean and standard deviation can be computed. The mean over the models is used as final prediction, while the standard deviation is used as an epistemic uncertainty measure. The descriptors used by MODNet are not necessarily numerically robust across the space of all possible compositions and crystal structures and thus could lead to unphysical results for particular rare representations (e.g., unit cells with very acute angles). To mitigate this, predictions from the final model that fall considerably outside of the range, $r=y_\mathrm{max}-y_\mathrm{min}$, of target values present in the training set (i.e., those outside of $r$ padded by 25\%) are replaced with a value drawn from a uniform distribution spanned by $r$. This makes the ensemble more robust; the remapping will have minor impact on the overall ensemble on average, but the variance will still be penalised with the number of pathological cases.

Second, MODNet has been made increasingly flexible in the nature of the material primitives supported (a full structure or just a composition) and on the types of target properties that can be learned. 
For instance, the user can manually choose which Matminer featurizers~\cite{WardMatminer} to apply to the data, or they can use presets suitable for structural or purely compositional samples.
Third, it is now possible to use MODNet to perform binary and multi-label classification tasks, predicting class membership probability in both cases.
Finally, an automated procedure for hyperparameter optimisation has been added within the framework of nested cross-validation (NCV).
Automated hyperparameter selection ensures a correct assessment of the generalisation error and minimises overfitting.
That is, hyperparameters should never be chosen to minimise a test error, but rather chosen internally with a validation procedure. 
The specific hyperparameter optimization process used in this work is described in \ref{appendix:hyperparameters}.

MODNet is an evolving framework and further improvements are planned.
We expect to apply it on even more datasets, as well as applying the trained models to accelerate materials design and discovery.
The source code and issue tracker, as well as several featurized datasets and pretrained benchmarks can be found at \href{https://github.com/ppdebreuck/modnet}{\texttt{ppdebreuck/modnet}} on GitHub~\cite{ModnetGithub}.

\section{Results and discussion}
\subsection{MatBench benchmarking}

The MatBench (v0.1) test suite contains 13 supervised ML tasks, taken from 10 datasets with size varying from 300 to 130,000.
Our previous benchmarking of MODNet has shown that it is most effective on small to medium-sized datasets, so efforts were focused on the 10 tasks with fewer than 15,000 samples. 
Of the remaining tasks, the largest two consist of the entire Materials Project~\cite{Jain2013}; the scaling of MODNet's performance on this dataset was previously studied in detail with the conclusion that graph-based networks outperform MODNet by a factor of 2 when the entire dataset is considered~\cite{debreuckMaterialsPropertyPrediction2021}.

The MatBench suite covers various observables in materials science driven by different length scales, from microscopic simulations of elastic, electronic, optical and vibrational properties, to macroscopic measurements of steel yield strengths and metallic glass formation.
For a detailed description of the contents and curation of each dataset, the reader is referred to the MatBench paper itself~\cite{dunnBenchmarkingMaterialsProperty2020}.

Each task provides either a composition or a structural model per sample.
The target is always a single property, either continuous or a discrete class label (such as metal or non-metal).
Almost all properties are therefore trained in single-target mode, i.e., one target per task.
The only exceptions are the elastic properties $\log_{10}K$ and $\log_{10}G$, where both bulk and shear modulus are provided for each sample (across two MatBench datasets); here model performance is improved by joint-learning both targets in a single model. Regression models were trained using the mean absolute error as a loss function.
For classification, the output activation is set to a softmax function and the categorical cross-entropy is used as loss function.

Our testing procedure follows the MatBench recommendations, i.e., an outer loop of five-fold cross-validation is used to determine the generalisation error (test error).
An inner five-fold validation is performed for both feature and hyperparameter selection, insuring an unbiased generalisation error. In other words, nested cross validation is used. More details of the hyperparameter optimisation process and training methods can be found in \ref{appendix:hyperparameters}.



\begin{table} 
\caption{
\label{tab:matbench_results}
The performance of MODNet (v0.1.9) on the MatBench (v0.1) test suite alongside alternative algorithms (Automatminer, RF, CGCNN and MEGNet) reported by Dunn \emph{et al.}~\cite{dunnBenchmarkingMaterialsProperty2020}.
The reported scores are either mean absolute errors (MAE) for regression or the area under the receiver-operator curve (ROC-AUC) for classification. 
These scores are averaged over the best models for each of the five cross-validation folds, following the recommendations of MatBench~\cite{dunnBenchmarkingMaterialsProperty2020}.
The hyperparameters are allowed to vary across these folds.
Numbers in bold indicate the best performing model for that task, and those within 2\% of the best performance relative to the best model.
A column is included for a null (``Dummy'') model that yields the mean of the dataset for each prediction for regression tasks, or returns the class frequency for classification tasks. } 
\begin{indented}
\scriptsize
\item[]\begin{tabular}{@{}lrllllll}
\br
MatBench task & $n$ & MODNet & AM & RF & CGCNN & MEGNet & Dummy\\
\mr
Steel yield strength (MPa) \cite{Citrine_steels} & 312 & \textbf{96.2}  & \textbf{95.2} & 104 & - & - & 230 \\
$E_\mathrm{exfol.}$ (meV/atom) \cite{choudharyHighthroughputIdentificationCharacterization2017} & 636 & \textbf{34.5} & 38.6 & 49.9 & 49.2 & 55.9 & 67.3 \\
$\mathrm{argmax}(\mathrm{PhDOS})$ (1/cm) \cite{petrettoHighthroughputDensityfunctionalPerturbation2018} & 1265 & 38.75 & 50.8 & 68 & 57.8 & \textbf{36.9} & 324 \\
Exp. band gap (eV) \cite{zhuoPredictingBandGaps2018}& 4604 & \textbf{0.347} & 0.416 & 0.446 & - & - & 1.14 \\
Refractive index \cite{petousisHighthroughputScreeningInorganic2017,Jain2013}& 4764 & \textbf{0.297} & \textbf{0.299} & 0.421 & 0.599 & 0.478 & 0.809 \\
Exp. metallicity \dag \cite{zhuoPredictingBandGaps2018} & 4921 & \textbf{0.970} & 0.92 & 0.917 & - & - & 0.495 \\
Glass-forming ability \dag \cite{LandoltBornstein1997, wardGeneralpurposeMachineLearning2016} & 5680 & \textbf{0.931} & 0.861 & 0.858 & - & - & 0.495 \\
$\log_{10}{K}$ ($\log_{10}{\mathrm{GPa}}$) \cite{dejongChartingCompleteElastic2015} & 10987 & \textbf{0.0548} & 0.0679 & 0.081 & 0.0712 & 0.0712 & 0.289 \\
$\log_{10}{G}$ ($\log_{10}{\mathrm{GPa}}$) \cite{dejongChartingCompleteElastic2015} & 10987 & \textbf{0.0731} & 0.0849 & 0.104 & 0.0895 & 0.0914 & 0.293 \\
Perovskite $E_\mathrm{form}$ (eV/atom) \cite{castelliNewCubicPerovskites2012} & 18928 & - & 0.194 & 0.235 & 0.0452 & \textbf{0.0417} & 0.566 \\
Band gap (eV) \cite{Jain2013}& 106113 & - & 0.282 & 0.345 & \textbf{0.228} & 0.235 & 1.33 \\
Metallicity \cite{Jain2013}& 106113 & - & 0.909 & 0.9 & 0.954 & \textbf{0.977} & 0.502 \\
$E_\mathrm{form}$ (eV/atom) \cite{Jain2013}& 132752 & - & 0.173 & 0.116 & 0.0332 & \textbf{0.0327} & 1.01 \\

\br
\end{tabular}
\end{indented}
\end{table}

\begin{figure}[ht]
    \centering
    \includegraphics[scale=1]{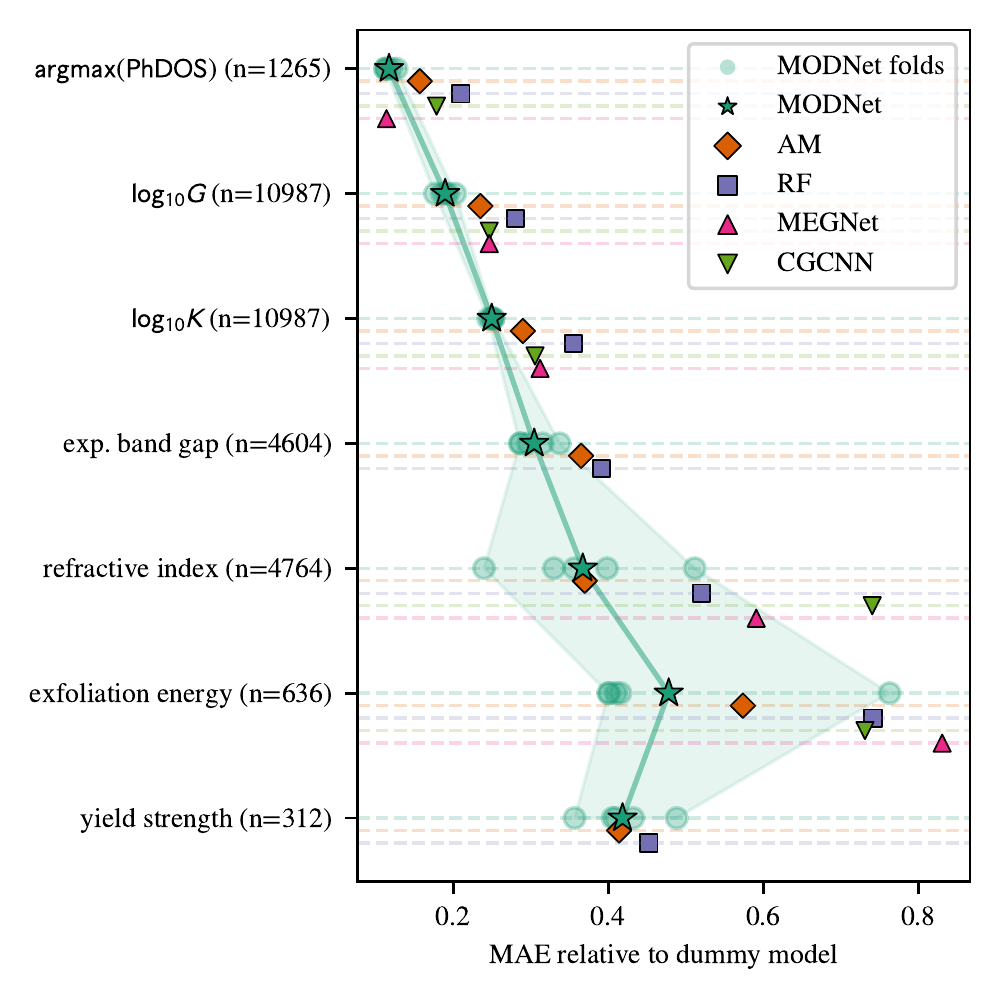}
    \caption{Comparison of model performance across regression tasks, relative to a dummy model that predicts the mean of the dataset. The shaded region indicates the range of errors spanned by the MODNet cross-validation folds. An arbitrary vertical offset has been applied to each model to aid visual discrimination.}
    \label{fig:regression_summary}
\end{figure}

Table~\ref{tab:matbench_results} contains an evaluation of MODNet on the MatBench suite, alongside five alternative algorithms reported on the MatBench leaderboard~\cite{dunnBenchmarkingMaterialsProperty2020}, as measured by the mean absolute error (MAE) for regression tasks, or the area under the reciver-operator curve (ROC-AUC) for classification tasks.
The Dummy algorithm forms a baseline by predicting the mean of the training set in case of regression, or a random label weighted by class size in case of classification.
A second baseline is formed by a Random Forest (RF) algorithm~\cite{breimanRandomForests2001} using Magpie elemental descriptors when only a composition is provided~\cite{wardGeneralpurposeMachineLearning2016}, and Sine Coulomb Matrix descriptor if structures are available~\cite{faberCrystalStructureRepresentations2015}.
MODNet is compared against Automatminer (AM) which is provided as a reference algorithm with the MatBench suite~\cite{dunnBenchmarkingMaterialsProperty2020}.
Automatminer creates automated end-to-end pipelines that couples materials datasets decorated with Matminer descriptors with an AutoML search using the TPOT package~\cite{OlsonGECCO2016}; this approach generates ensembles of tree-based models to perform the final prediction.
It has the advantage of being fully automated and therefore requires minimal ML expertise to run on a new dataset.
Finally, for tasks that contain the crystal structure as input, the deep learning models CGCNN~\cite{xieCrystalGraphConvolutional2018} and MEGNet~\cite{chenGraphNetworksUniversal2019} are also compared against; these models perform convolution and pooling operations on a graph representation of the crystal structure, allowing for structurally local interpretations of predicted properties.

The presented results show that MODNet significantly improves the error on 4 tasks (exp. band gap, exp. metallicity, glass-forming ability, 2D exfoliation energy), and matches/outperforms the existing leader (within 2\%) error for 3 tasks, and finally has a higher error on only one task (steel yield strength). 
Figure \ref{fig:regression_summary} displays the results of the regression tasks relative to the Dummy model across all datasets. 
One can see that the Dummy algorithm is outperformed by all models by a factor of two to three. The spread between the performance on individual folds is correlated both with the improvement relative to the Dummy model, but also to the spread between all competing approaches.
It should be noted that hand-tuned models after automatic hyperparameter optimisation can also lower this error further.
This effect was most prominent for the steel yield strength task where model performance could be improved by nearly 20\% with manual tuning, indicating that the hyperparameter optimisation approach could be improved for smaller datasets where a simple grid search is inherently noisier.

Detailed MODNet benchmarking plots for each dataset can be found in \ref{appendix:benchmarking}. These plots display the typical regression between target and predicted properties, and additionally a regression between targets and prediction error to show any model biases as a function of target property. As one might expect, the sample density is often considerably lower for extremal target property values, so many datasets display a positive correlation between the errors and targets (i.e., predictions decrease in quality for larger target values). This effect is most prominent for the refractive index and exfoliation energy datasets shown in Figures \ref{fig:matbench_dielectric} and \ref{fig:matbench_jdft2d}.



As a general trend, it can be observed that MODNet performs relatively well on tasks that are limited to composition data only.
We explain this success by the usage of a limited input neural network where a latent representation is learned for the different elemental contributions; and moreover acts as a dimensionality reduction technique. This representation is much more flexible than the equivalent obtained by feature engineering for tree-based approaches.
Recently presented deep representation learning approaches should be able to exploit this effect even further, especially if transfer learning is used~\cite{goodall2020predicting, wang2020compositionally}.




\subsection{Model quality and uncertainty assessment}


When predicting material properties with a machine learning model, one should keep in mind that the prediction error can vary significantly from material to material.
The MAE is merely an aggregate statistic, with sometimes huge differences (lower or higher) on single predictions.
For instance, concerning the experimental band gap, the errors span 5 orders of magnitude from $10^{-4}$ to $10^1$.
This has two major consequences.
First, a more comprehensive way of assessing a model's performance is needed.
Particularly, when comparing different models, it is generally worth having a compilation of metrics or a full distribution of errors instead of relying on an individual metric.
Second, the ability to provide an error estimate on predictions \emph{a priori} is extremely useful, so one could be tempted to seek to quantify this uncertainty.
In other words, given the input material, how reliable is the final prediction? This section covers a solution to the first question, and proposes a simple distance metric in feature space to address the second question.

\subsubsection{Model quality assessment}
The results in Table~\ref{tab:matbench_results} are an average over the 5 cross-validation folds, and significant variations between folds are often observed, as shown in Figure \ref{fig:regression_summary}.
This is caused by the underlying spread in errors over the individual predictions, and when sampled, a strong discrepancy between folds can be seen, especially when the dataset is small.
This variance over the points contains valuable information about the intrinsic model performance, and should, in our opinion, not be neglected.
Too often in the materials science field, models are assessed using a single metric, which is in general not enough to fully capture and compare ML models.
Two models could have a similar MAE, but with a very different spread.
Moreover, models can set their internal parameters to suit the benchmark metric (e.g., by setting the loss function the the metric of interest), and therefore optimise their model with respect to a benchmark rather than the domain of application. 
A more holistic approach to compare models is clearly needed. 
The advantage of using different metrics is that they illuminate different performance aspects of a model, and taken together, they give a more comprehensive view of a model's quality.
Following the work of Vishwakarma~\textit{et al.}~\cite{vishwakarmaMetricsBenchmarkingUncertainty2020}, we provide a compilation of metrics to assess MODNet performance.

We suggest the following metrics for regression: mean absolute error (MAE), median absolute error (MedAE), root-mean-squared error (RMSE), mean absolute percentage error (MAPE), maximum absolute error (MaxAE), and the Pearson correlation ($R$) of an ordinary least squares regression between the target and predicted properties. The corresponding values for the different tasks on MODNet are given in Table~\ref{tab:matbench_metrics} (excluding MAPE).

Comparing the MAE with the RMSE offers insights on the variability of prediction errors.
Large fluctuations in errors (outliers) will result in a significantly higher RMSE.
All tasks in Table~\ref{tab:matbench_metrics} have a higher RMSE than MAE, indicating a right skewed distribution of the absolute errors. This is especially the case for the refractive index and exfoliation energy.

The MAPE provides an error \emph{relative} to the ground truth, and complements thus the MAE and RMSE.
It should be noted, however, that it is best avoided to properties having values equal or close to zero, as the MAPE diverges.
It is therefore omitted in Table~\ref{tab:matbench_metrics}.

The MaxAE gives the worst error in the test set, and is thus related to the spread in errors.
When making finite-cost decisions based on predictions from a model, it is crucial to have an idea of the worst case error.
As can be seen from the table, all tasks have a MaxAE that is orders of magnitude larger than the MAE among the test samples. 
This aspect and its relation to uncertainty is discussed hereafter. 

The $R$ value measures how close predictions are to the ground truth by measuring their linear relationship, and has the advantage to be bounded between -1 and 1. A value of 1 indicates a perfect positive correlation. This enables to compare performance not only between models but also between tasks. Best performing tasks are found to be the phonons and elastic constants. In contrast, the refractive index and exfoliation energy are both seen to yield a low $R$ (0.43 and 0.65 respectively). This is mainly caused by outliers, as the $R$-value metric is very sensitive to them.

Finally, an alternative solution to the ensemble of metrics, is to provide a full probability distribution of errors, by for instance applying a Gaussian kernel density estimation on the discrete errors.
An example is given in the previous work on MODNet~\cite{debreuckMaterialsPropertyPrediction2021}.

Concerning classification, the previously mentioned ROC-AUC is an overall good metric for binary classification provided the labels are balanced, combining the model's performance at different class probability thresholds.
Providing the full ROC curve is also good practice.
However, for imbalanced datasets, precision-recall (PR) curves have greater utility~\cite{davisRelationshipPrecisionRecallROC2006}. Both the ROC and PR curves are given in \ref{appendix:benchmarking} for the experimental metallicity and glass-forming ability tasks using MODNet. The corresponding ROC-AUC and average precision scores are given in Table~\ref{tab:matbench_metrics}.
For multi-class classification, a confusion matrix is a simple yet powerful way of visualizing the quality of the model for a given threshold.






\begin{table} 
\caption{\label{tab:matbench_metrics}
A battery of metrics indicating MODNet performance on regression and classification tasks of MatBench v0.1. $\dagger$ As in Table \ref{tab:matbench_results}, one numerically unstable prediction was replaced with the dataset mean before reporting the metrics for the $\log_{10}K$ and $\log_{10}G$ tasks.}

\begin{indented}
\scriptsize
\item[]\begin{tabular}{@{}lrrrrr}
\br
Regression Task & MAE & Median AE & RMSE & MaxAE & $R$ \\
\mr
Steel yield strength (MPa) & 96.2 & 60.8 & 151.9 & 931 & 0.87 \\
$E_\mathrm{exfol.}$ (eV/atom) & 34.5 & 9.05 & 102.8 & 1535 & 0.65 \\
Refractive index & 0.297 & 0.0612 & 1.90 & 58.9 & 0.43 \\
Exp. band gap (eV) & 0.347 & 0.080 & 0.75 & 9.9 & 0.86 \\
$\log_{10}{K}$ ($\log_{10}{\mathrm{GPa}}$)$\dagger$ & 0.0548 & 0.028 & 0.104 & 1.5 & 0.96 \\
$\log_{10}{G}$ ($\log_{10}{\mathrm{GPa}}$)$\dagger$ & 0.0731 & 0.049 & 0.110 & 1.7 & 0.96 \\
$\mathrm{argmax}(\mathrm{PhDOS})$ (1/cm) & 38.75 & 20.7 & 79.4 & 1032 & 0.99 \\
\end{tabular}

\item[]\begin{tabular}{@{}lrr}
\br
Classification Task & ROC-AUC & Average Precision \\
\mr
Exp. metallicity & 0.970 & 0.968 \\
Glass-forming ability & 0.931 & 0.966 \\
\end{tabular}
\end{indented}
\end{table}

\subsubsection{Uncertainty prediction: the bias-imbalance issue}
\label{sec:uncertainty}

Beyond benchmarking a model, it is equally important to asses the uncertainty of novel predictions.
Given an unseen sample, how reliable is the model's prediction? This aspect is closely related to the applicability and generalisation capabilities of a model.
This is not a straightforward problem in materials science and it should be given significant attention, in the same way much effort is spent on optimizing an error metric.
For instance, a MODNet model was created on the DFT-predicted refractive index dataset of Naccarato~\textit{et al.}~\cite{naccaratoSearchingMaterialsHigh2019}.
A random test set of 200 samples yields an MAE of 0.051.
However, when looking only at the non-oxides in this test set, the error jumps to 0.082.
The oxides (defined loosely here as compounds containing at least one oxygen atom) in the test set have a much lower MAE of 0.035.
Similarly, for the bulk modulus in MatBench, we observed an error of over 50,000 $\log_{10}$(GPa) on a particular compound, as noted in the caption of Table \ref{tab:matbench_results}.
These variations are significant and anticipating such fluctuations \emph{a priori} is therefore important for reliable real-world applications.


From physical and chemical grounds, we learn that materials appear in distinct qualitative classes based on composition (e.g., oxides, phosphides, sulfides), structural classes (e.g., perovskite, wurtzite, zincblende) or by intrinsic properties (e.g., metals and non-metals).
Other more subtle possibilities exists and are equally important as materials often behave differently depending on subgroup membership.
Therefore, it could be expected that a machine learning model trained on a particular class will fail to make good predictions when applied to another class.
However, many of the different classes mentioned previously are principally human made. They are based on human intuition and concepts such as bond type or the periodic table, and in some sense, do not exist for the machine~\cite{georgeChemistMachineTraditional2020a}.
In all generality, a machine learning model goes beyond these human concepts and forms its own material representation, considering a material as a purely mathematical object in a high-dimensional space.
For instance, one can use the feature space, or better, a condensed latent space.
In this work, we consider materials either as points in the MODNet feature space, or as a PCA (principal component analysis) reduction in a lower dimensional space.
Classes can now appear naturally as point clouds in these spaces and a similarity metric can be defined through a simple distance-based measure.
Close materials are similar and should behave closely with respect to the learned property; if this is not the case, the feature space may not be sufficiently complete to fully discriminate samples.

The concepts of \emph{imbalance} and \emph{bias} are important in this context.
A training set that has two or more distinct classes (i.e., clusters in the material space) that are sampled in different proportions is \emph{imbalanced}.
For instance, 84\% of the refractive index dataset is comprised of oxides and therefore constitutes an imbalanced training set.
A training set is \emph{biased} when it covers only a specific region (i.e., not all classes are covered) of the material space, a more extreme case of imbalance.
Both imbalance and bias have consequences to the generalisation of a model as they express potential dissimilarities between training and target domain.

\begin{figure}[t]
\caption{PCA decomposition of the refractive index dataset (Naccarato~\textit{et al.}~\cite{naccaratoSearchingMaterialsHigh2019}) for the second and third component.
The second component is linked to the bond-length, while the third component is linked to the ionicity of the compound.
Data points corresponding to oxides and non-oxides are coloured blue and red respectively.
Compounds having at least one oxygen atom tend to have shorter mean bond lengths and increased ionicity.
An imbalance appears due to sampling: the dotted blue line divides the sample space in denser and coarser regions.
The compounds in the well-sampled centre region yield on average the lowest error, regardless of whether they are an oxide.
PCA components are provided in detail in \ref{appendix:pca}.}
\label{fig:pca}
\centering
\includegraphics{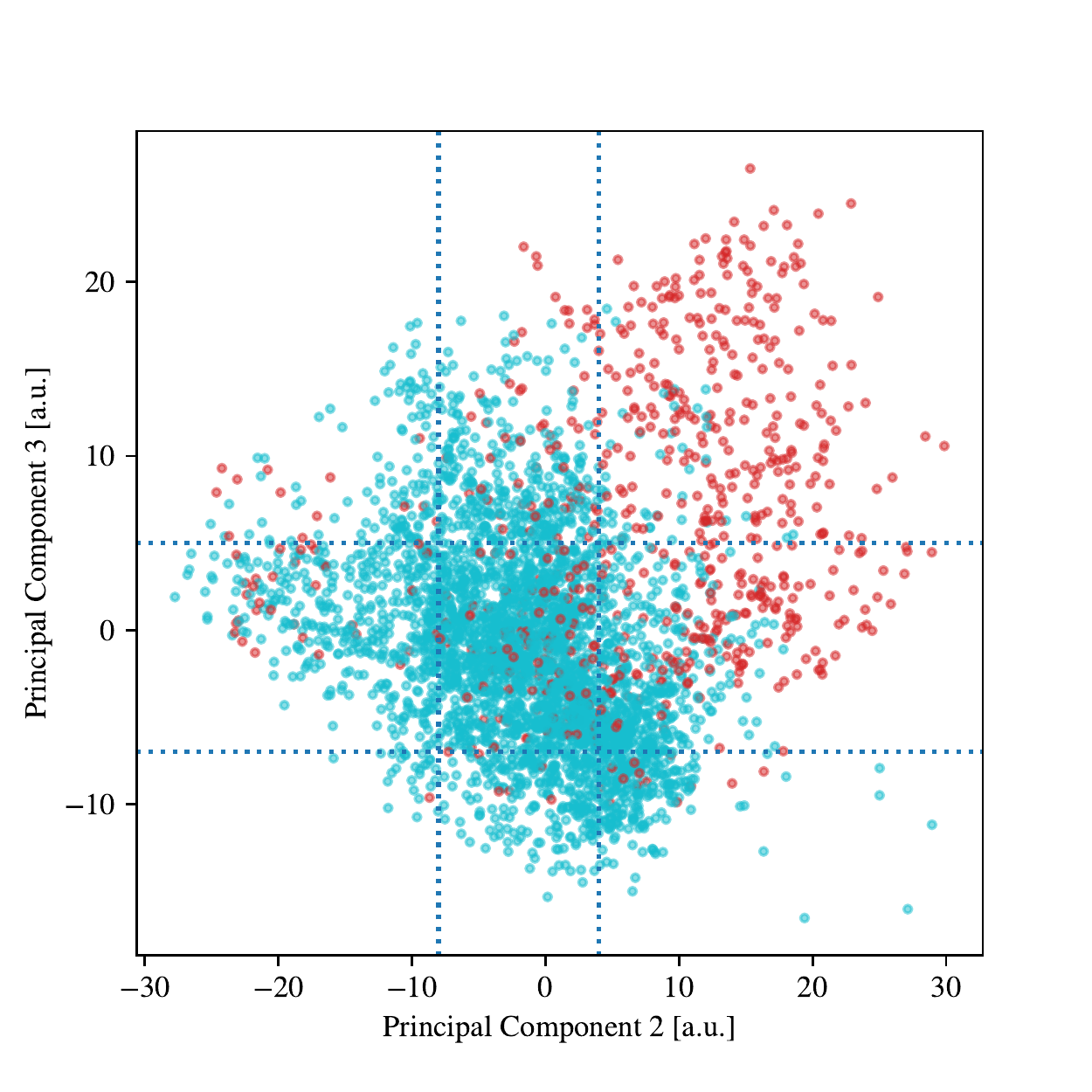}
\end{figure}

In order to visualise imbalance-bias and the discrepancy in error between oxides and non-oxides for the refractive index DFT dataset of Naccarato~\textit{et al.}~\cite{naccaratoSearchingMaterialsHigh2019}, we performed a PCA decomposition.
The first three components together account for 25 \% of the variance.
Their detailed description are provided in \ref{appendix:pca}, with the list of descriptors and corresponding weights in the feature space.
The first feature roughly corresponds to a mean of all features and therefore identifies outlier compounds in the feature space.
They are not of particular interest in our study case.
However, the second and third components are closely related to the concept of oxide.
Figure~\ref{fig:pca} displays the second and third principal components of the PCA decomposition.
Each data point, initially a high-dimensional point in the MODNet feature space, is now projected onto two dimensions.
The second component is proportional to the mean bond length (or mean atomic volume), with corresponding descriptors of radial distribution function peaks and elemental radii.
The third component is related to the bond type, comprising of descriptors related to electronegativity and element identity.
Small values indicate structures tendentiously forming ionic bonds, while larger values indicate metallic bonding.
The second component can be seen as being related to the structure while the third depends more on its composition.
From the figure, it can be seen that regions with dense (lower left) and coarse (upper right) samplings appear, indicating an imbalance in the dataset.
Notably, oxides (in blue) are on average in the lower left part and non-oxides are on average in the upper right corner of the decomposition.
Indeed, oxygen being a quite small and electronegative atom, it tends to shorten the mean bond length of a compound and increases ionicity.
These two observations explain the higher average error on non-oxides: they form a minority group in a region with a coarse sampling, with fewer neighbouring samples than oxides.
Importantly, it is the density (thus the number of neighbours) that is the critical parameter here.
Therefore, it is possible to have non-oxides that are well predicted and oxides that are poorly predicted; non-oxides in the middle region of Figure~\ref{fig:pca} have a MAE of 0.04, while oxides in the lower right region have an error of 2.5.
Note that this latter region is mainly outside the main distribution of points, and corresponds thus to a biased training set.

The oxide vs. non-oxide distinction is used here only as an illustrative example. We emphasize that the imbalance should be seen in the feature space directly, i.e., in terms of bond-length / bond-type imbalance.

\begin{figure}[h]
\caption{Confidence error curves for six different regressions task found in MatBench: steel yield strength, 2D exfoliation energy, refractive index, experimental band gap, phonon DOS peak and bulk modulus. Each curve represents how the mean absolute error changes when test points are sequentially removed following different strategies. A baseline (red) is generated by randomly removing points one at a time (as if all points had equal confidence), with the shaded area showing the deviation across 1000 trials. The randomly ranked error (red) forms a baseline with the shaded area representing the standard deviation over 1000 random runs. The error ranked curve (green), where the highest error is removed sequentially, represent a lower limit. The std-ranked strategy follows the uncertainty predicted by the ensemble MODNet, while the $d_\mathrm{KNN}$ ranking is based on the 5-nearest neighbour cosine distance between each test point and training set.}
\label{fig:ranked_mae}
\centering
\includegraphics[width=\textwidth]{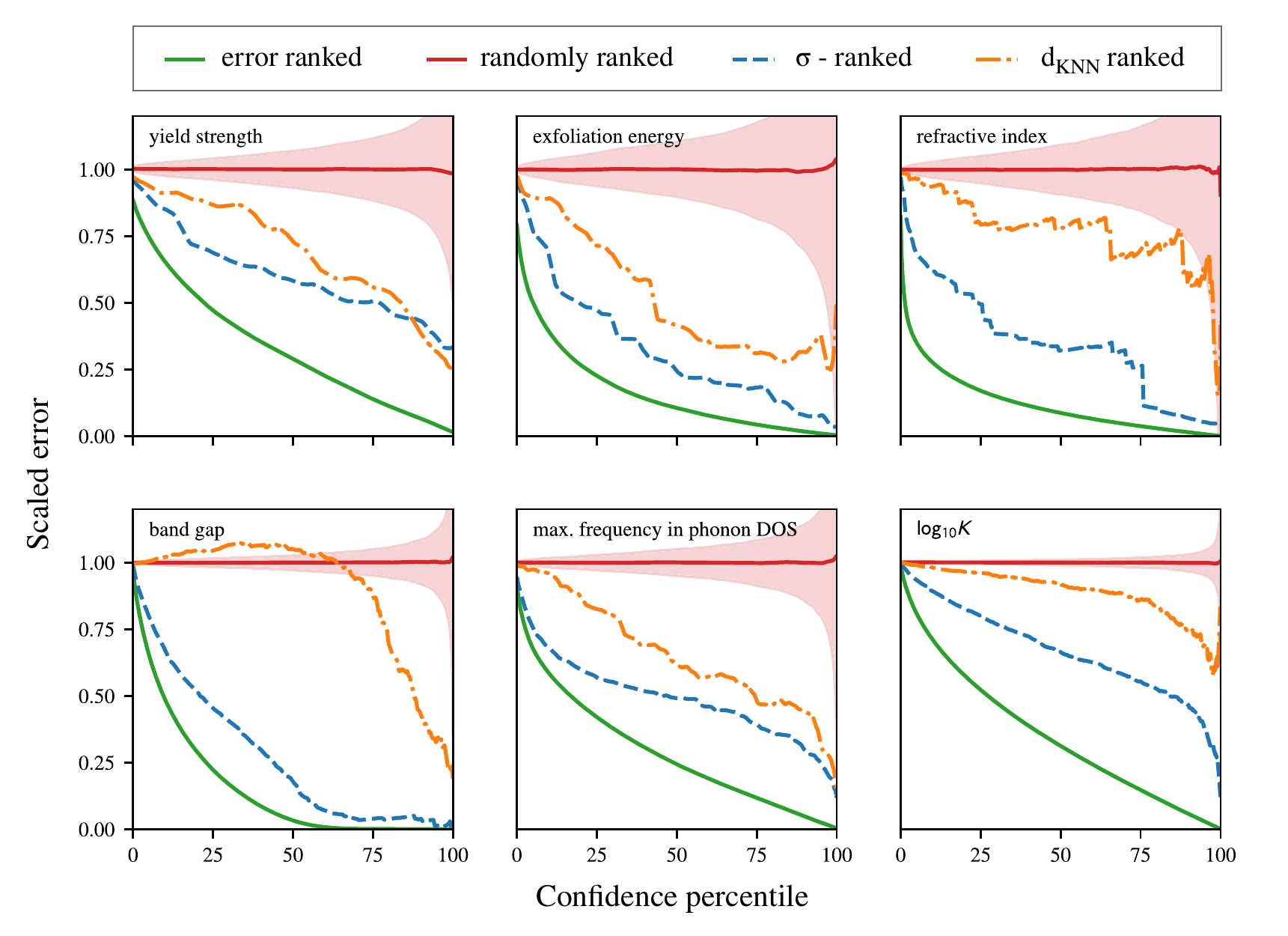}
\end{figure}

In order to have a more quantitative analysis on the uncertainty, the standard deviation provided by the ensemble method is used. The ensemble method includes 125 MODNet models with varying architecture, initial weights and bootstrapped training set (see Methods). The predicted variance therefore only accounts for epistemic uncertainty~\cite{lakshminarayananSimpleScalablePredictive2017,scaliaEvaluatingScalableUncertainty2020}.

Figure~\ref{fig:ranked_mae} represents the MAE as a function of the confidence percentile for different MatBench tasks. The MAE of a confidence percentile $p$ is computed by removing the $p$\% most uncertain predictions, following different strategies. Randomly removing points (in red) forms a baseline, while a perfect ranked strategy based on the absolute errors forms a lower limit. The uncertainty provided by the ensemble MODNet model is depicted in blue. In addition, a strategy based on distances in the feature space is added. The idea is to determine a model's applicability domain via the distance of a query point and the $k$-nearest neighbours (KNN) in the training set, $d_\mathrm{KNN}$. The space of optimal descriptors, as selected by MODNet is used. Here, after various tests, the cosine distance is used for simplicity with $k=5$. Other distances are possible, although one should be cautious with the high dimensionality~\cite{aggarwalSurprisingBehaviorDistance2001}.

Both the ensemble's uncertainty and $d_\mathrm{KNN}$ distance are adding value to distinguish error magnitudes with respect to the random procedure. The ensembling techniques is seen to be performing particularly well, especially at the lower end of the confidence percentiles. Moreover, assuming Gaussian distributions, 95\% confidence intervals can be built for each prediction from the predicted standard deviation. Figures
in~\ref{appendix:benchmarking} depict confidence intervals for the different predictions as well as calibration curves, and various error related scatter plots. In particular, good agreement between predicted and observed test-error distribution are found. This shows that the ensemble MODNet model is effective at providing uncertainty estimates. Note that recalibration of the model (i.e., remapping the predicted standard deviations based on the observed errors) could further improve the accuracy.

The improvement of $d_\mathrm{KNN}$ wrt. random ranking confirms that feature space imbalance is indeed an important factor in assessing uncertainty. However, it has clear limitations. First, it is clear that it does not perform as well as the probabilistic ensemble method. Especially, for the experimental band gaps a poor performance is found. Second, the $d_\mathrm{KNN}$ method cannot provide any quantitative uncertainty estimate such as confidence intervals. Therefore, the ensembling technique is preferred for uncertainty estimations.

Finally, it should be noted that beyond ensemble-based methods, there are other model architectures that intrinsically perform uncertainty quantification, namely Bayesian learning approaches, Gaussian processes or random forests~\cite{abdarReviewUncertaintyQuantification2021, coulstonApproximatingPredictionUncertainty2016}.

We encourage similar analysis on other models.
It will enable both model applicability assessment (comparison between different models), and uncertainty estimation for a given prediction (within a single model), without incurring a great computational or architectural cost.
If robustly identified, the latter contains valuable information about the confidence for new predictions.

\section{Summary}

To summarise, we report benchmarks of the Materials Optimal Descriptor Network (MODNet) on the MatBench v0.1 test suite.
MODNet is a universal model, in regard to both the inputs (material structure or composition) and the target type.
MatBench is a standardised benchmarking pipeline, with fixed procedures on training and testing, containing various material prediction tasks of differing sizes and types.
Our results show that MODNet performs well on small to medium-sized datasets.
In particular, it was found to outperform or match the current leaders on the experimental band gap, experimental metallicity, glass formation ability, 2D exfoliation energy, elastic moduli (bulk and shear), and phonon DOS peak estimation (see Table~\ref{tab:matbench_results}).
We provide metrics beyond MAE and ROC-AUC scores and encourage the reporting of multiple metrics and we make available all benchmarking data for future model comparisons.

We show that depending on the test set sampling (or fold), significant variation in error can be measured.
This is due to the high dimensionality of the feature space spanned by materials and the associated sparsity of a given dataset.
Bias or imbalance is easily introduced in the training set by this sampling.
Therefore, applicability through uncertainty assessment is a crucial aspect for future developments.
We show that dataset imbalance can be examined through dimensionality reduction techniques (such as PCA) and that uncertainty can be quantified by ensemble-based methods.
This offers the possibility to set an confidence bound on individual predictions; uncertain predictions could be flagged or removed.

Finally, we emphasise that this work only forms a basis to a better practice in model design and performance assessment.
Uncertainty quantification, and particularly for extrapolations, is a difficult task in general.
We showed some options to this regard, but further development and benchmarking is necessary.

We hope that this work encourages future developments on the topics of metrics, bias and uncertainty that are crucial in materials science for robust, transparent testing and better understanding of a model applicability in a wider context.
By providing all of our benchmarking data, we hope that the differences between competing approaches can be learned from and used to improve real-world model performance.


\section*{Acknowledgements}
P.-P.~D.B. and G.-M.~R. are grateful to the F.R.S.-FNRS for financial support.
M.~L.~E. and G.-M.~R. acknowledge support from the European Union's Horizon 2020 research and innovation program under the European Union's Grant agreement No. 951786 (NOMAD CoE).

Computational resources were provided by the supercomputing facilities of the Université catholique de Louvain (CISM/UCL) and the Consortium des Équipements de Calcul Intensif en Fédération Wallonie Bruxelles (CÉCI) funded by the Fond de la Recherche Scientifique de Belgique (F.R.S.-FNRS) under convention 2.5020.11 and by the Walloon Region.

\section*{References}
\bibliography{main.bib}
\clearpage

\appendix

\section{The MODNet model}
\label{appendix:MODNet}
This appendix briefly summarizes the MODNet model. For further details, the reader is referred to Ref.~\cite{debreuckMaterialsPropertyPrediction2021}.

\begin{figure}[!htb]
\centering
\includegraphics{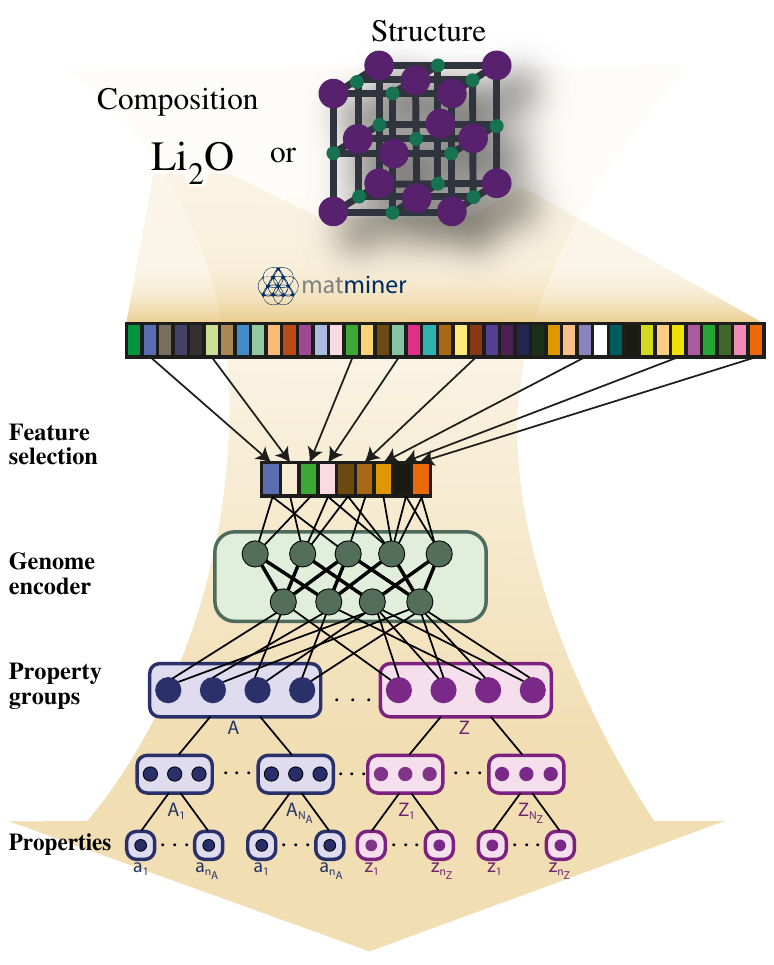}
\caption{
Schematic of the MODNet model (adapted from Ref.~\cite{debreuckMaterialsPropertyPrediction2021}).
The feature selection on matminer (from structure or composition) is followed by a hierarchical tree-like neural network. Various properties $A_1$,\ldots,$A_{N_A}$,\ldots,$Z_1$,\ldots,$Z_{N_Z}$ (e.g. Young's modulus, refractive index, ...) are gathered in groups from $A$ to $Z$ of similar nature (e.g. mechanical, optical, ...). Each of these may depend on a parameter (e.g. temperature, pressure, ...): $A(a)$,\ldots,$Z(z)$. The properties are available for various values of the parameters $a_1$,\ldots,$a_{n_A}$,\ldots,$z_1$,\ldots,$z_{n_Z}$. The
first green block of the neural network encodes a material in an appropriate all-round vector, while subsequent blocks decode and re-encode this representation in a more target specific nature.
\label{fig:model_schematic}}
\end{figure}

The model, named Material Optimal Descriptor Network (MODNet) and illustrated in Figure~\ref{fig:model_schematic}, consists in  a feedforward neural network with an reduced set of optimal descriptors (based on a criterion detailed further). Moreover, we propose an architecture that, if desired, learns on multiple properties, with good accuracy. This makes it easy to predict more complex objects such as temperature-, pressure-, or energy-dependent functions.

The structure (or composition) is first transformed to descriptors based on physical, chemical, and geometrical properties. They fulfill a number of constraints such as rotational, translational and permutational invariances. Moreover, features driven by physical and chemical intrinsically contain knowledge (wrt. more flexible graph representations) that facilitates learning when dealing with limited datasets. We rely on a large amount of features previously published in the literature, that were centralized into the matminer project~\cite{wardMatminerOpenSource2018}. In order to reduce redundancy and therefore limit the curse of dimensionality~\cite{jovicReviewFeatureSelection2015} we rely on a feature selection process based on the \emph{Normalized Mutual Information} (NMI). It is defined as,
\begin{equation}
\label{eq:NMI}
    \mathrm{NMI}(X,Y) = \frac{\mathrm{MI}(X,Y)}{(\mathrm{H}(X)+\mathrm{H}(Y))/2}
\end{equation}
with MI the {mutual information}, computed as described in Ref.~\cite{kraskovEstimatingMutualInformation2004} and H the \emph{information entropy} ($\mathrm{H}(X)=\mathrm{MI}(X,X)$). The NMI, which is bounded between 0 and 1, provides a measure of any relation between two random variables X and Y. It goes beyond the Pearson correlation, which is parametric (it makes the hypothesis of a linear model) and very sensitive to outliers.

Given a set of features $\mathcal{F}$, the selection process for extracting the subset $\mathcal{F}_S$ goes as follows. When the latter is empty, the first chosen feature will be the one having the highest NMI with the target variable $y$. Once $\mathcal{F}_S$ is non-empty, the next chosen feature $f$ is selected as having the highest relevance and redundancy (RR) score:
\begin{equation}
    \mathrm{RR}(f) = \frac{\mathrm{NMI}(f,y)}{\Big[\max_{f_s \in \mathcal{F}_S}\big(\mathrm{NMI}(f,f_s)\big)\Big]^{p}+c}
    \label{eq:RR}
\end{equation}
where $(p,c)$ are two hyperparameters determining the balance between relevance and redundancy. In practice, varying these two parameters dynamically seems to work better, as redundancy is a bigger issue with a small amount of features. 

The selection proceeds until the number of features reaches a threshold which can be fixed arbitrarily or, better, optimized such that the model error is minimized. When dealing with multiple properties, the union of relevant features over all targets is taken. 


Secondly, we take the advantage of learning on multiple properties simultaneously.
This could be used, for instance, to predict temperature-curves for a particular property.

In order to do so, we use the architecture presented in Figure~\ref{fig:model_schematic}. Here, the neural network consists of successive blocks (each composed of a succession of fully connected and batch normalization layers) that split on the different properties depending on their similarity, in a tree-like architecture. The successive layers decode and encode the representation from general (genome encoder) to very specific (individual properties). Layers closer to the input are shared by more properties and are thus optimized on a larger set of samples, imitating a virtually larger dataset. These first layers gather knowledge from multiple properties, known as joint-transfer learning~\cite{liLearningForgetting2018}. This limits overfitting and slightly improves accuracy compared to single target prediction.

\newpage
\section{Notes on featurization and training}
\label{appendix:hyperparameters}

All datasets were featurized using the \texttt{DeBreuck2020Featurizer} preset that is bundled with MODNet, though some datasets only made use of compositional descriptors. After featurization, the normalised mutual information was computed between all pairs of descriptors across the entire dataset and was referred back to when training each model.

\begin{figure}[h]
    \centering
    \includegraphics[scale=1]{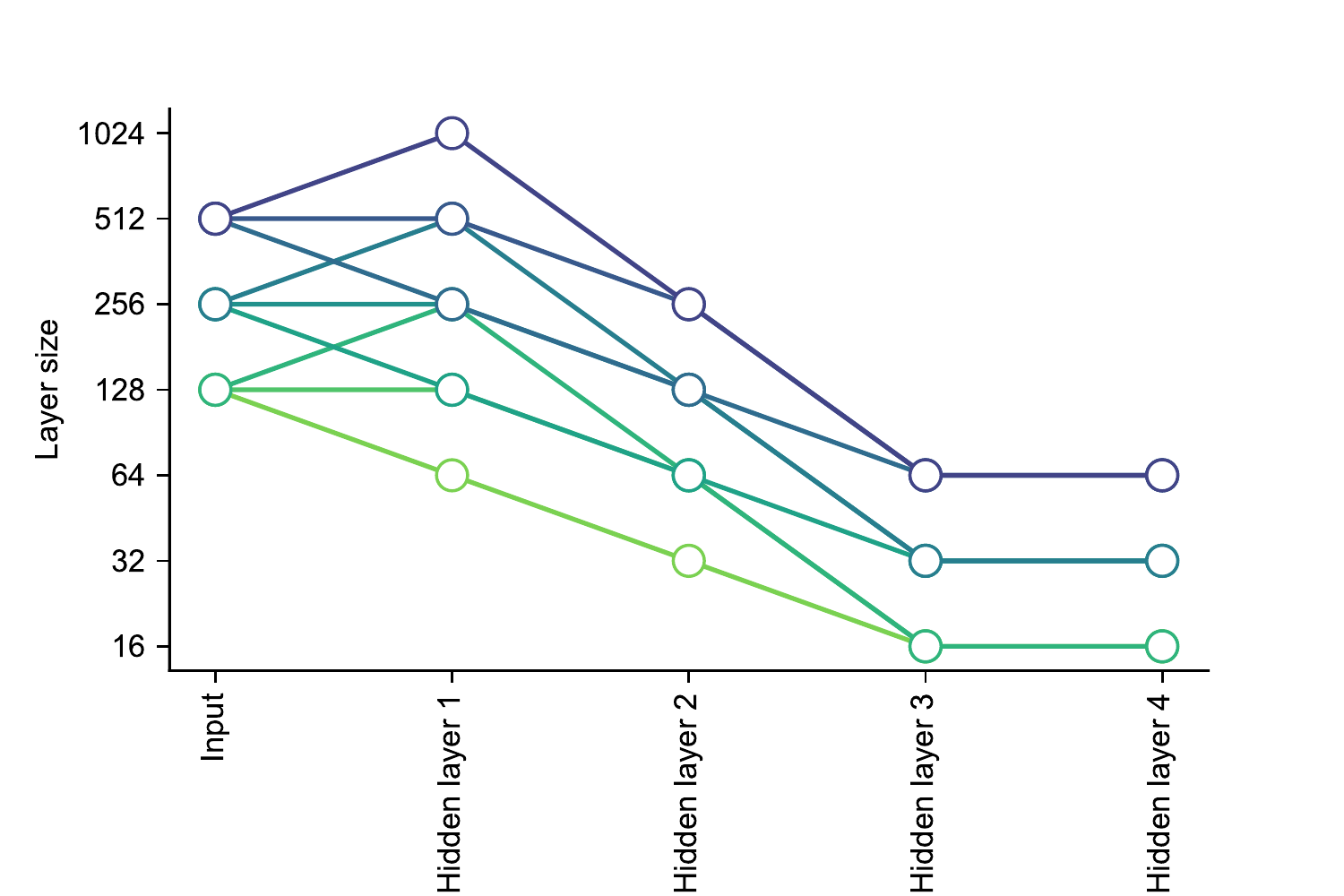}
    \caption{Graphical depiction of the network architectures sampled during hyperparameter optimisation. Absolute values correspond to the layer depths for a 1000-dimensional feature space (before feature selection).}
    \label{fig:network_hyperparameters}
\end{figure}

Hyperparameter optimization was performed via 5-fold cross-validation, with 85\% of each fold being used for training in the inner loop. For each fold, a grid search over batch sizes, learning rates, number of training features ($N$) and hidden layer depths was performed, and the hyperparameters with the lowest MAE on the remaining 15\% of the fold were used to fit a new model on the entire fold. Feature selection is key to the performance of MODNet; the features were selected per fold based on a relevance-redundancy criterion and the top $N$ features were used for training. Some typical architectures are illustrated in Figure~\ref{fig:network_hyperparameters} for a 1000-dimensional feature space. 

The outer loop was performed in parallel, yielding five models per task. After featurization, training can be efficiently performed on commodity hardware, with a full grid search (20-60 hyperparameter combinations) requiring no more than a 2 hours per outer fold when run on 4-cores of an AMD EPYC 7742 (64-core) CPU.

All benchmarking data and the scripts used to featurize and train models for each dataset are available on GitHub at \href{https://github.com/ml-evs/modnet-matbench}{\texttt{ml-evs/modnet-matbench}}.


\clearpage
\section{Detailed benchmark plots}
\label{appendix:benchmarking}


\subsection{\texttt{matbench\_dielectric}}
\begin{figure}[h]
    \centering
    \includegraphics[width=0.495\textwidth]{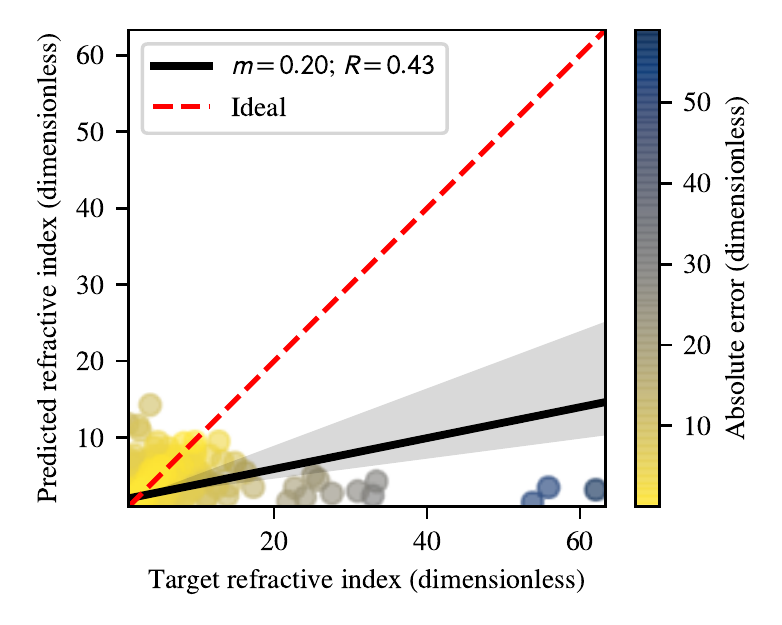}
    \includegraphics[width=0.495\textwidth]{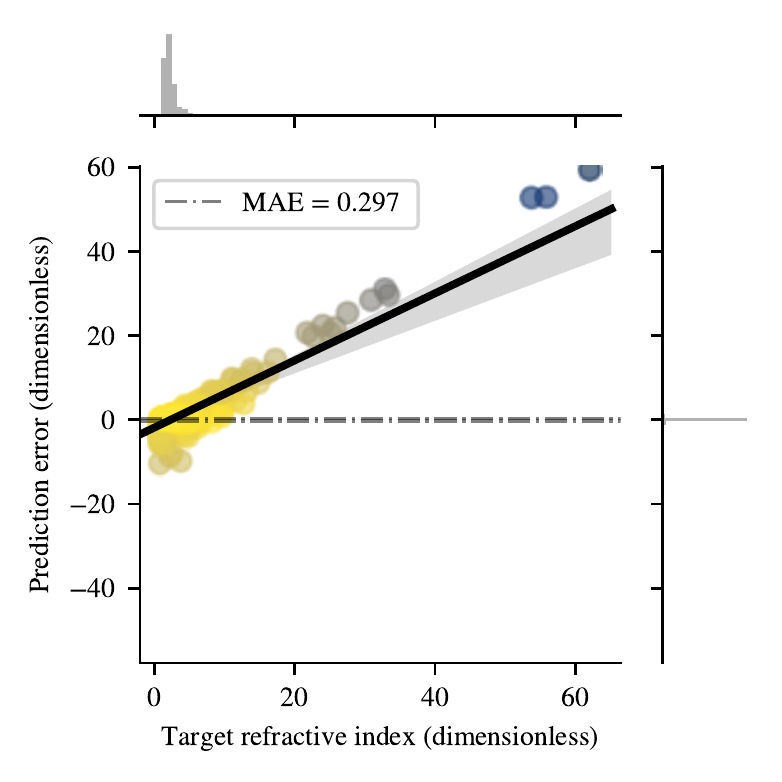}
    \caption{Results for the \texttt{matbench\_dielectric} dataset.}
    \label{fig:matbench_dielectric}
\end{figure}

\begin{figure}[h]
    \centering
    \includegraphics[width=\textwidth]{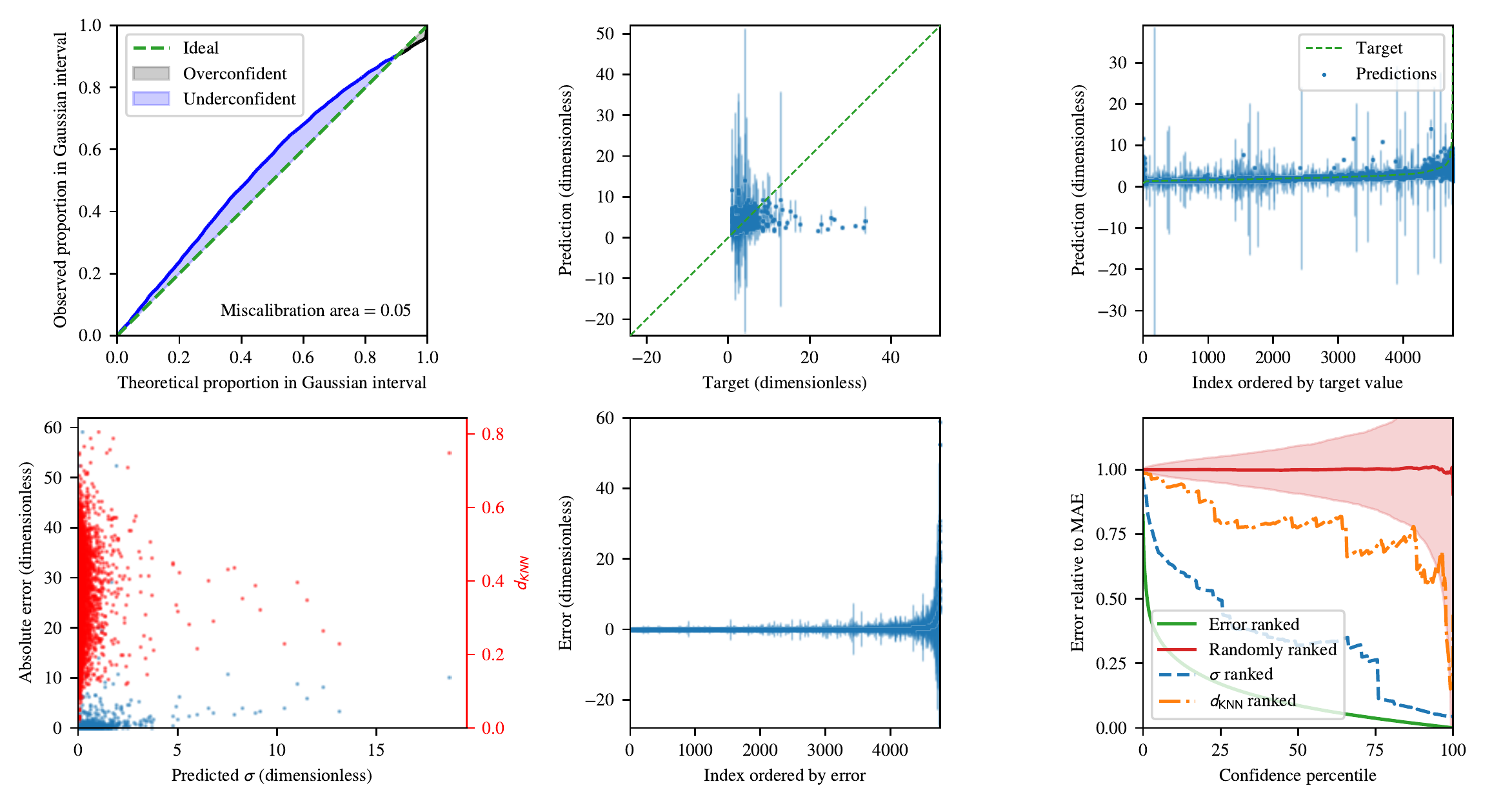}
    \caption{Uncertainty results for the \texttt{matbench\_dielectric} dataset.}
    \label{fig:unc_dielectric}
\end{figure}

\clearpage
\subsection{\texttt{matbench\_expt\_gap}}

\begin{figure}[h]
    \centering
    \includegraphics[width=0.495\textwidth]{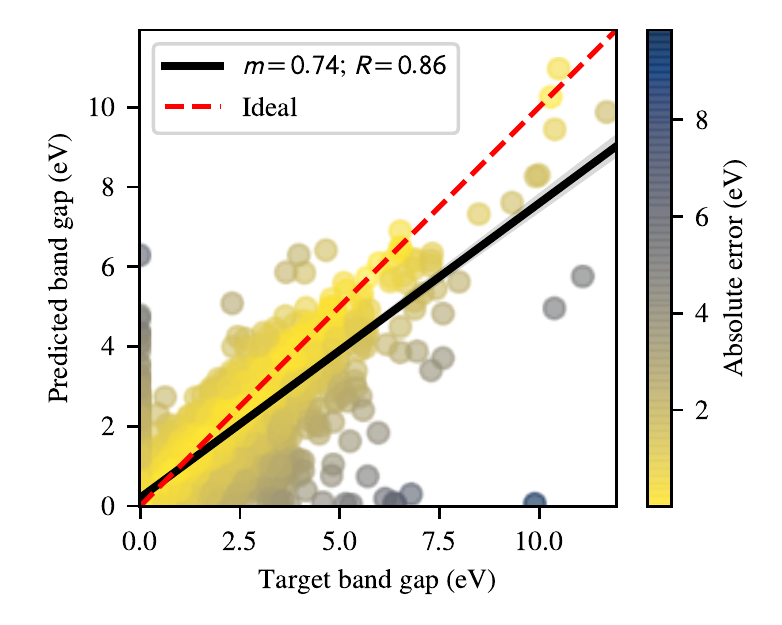}
    \includegraphics[width=0.495\textwidth]{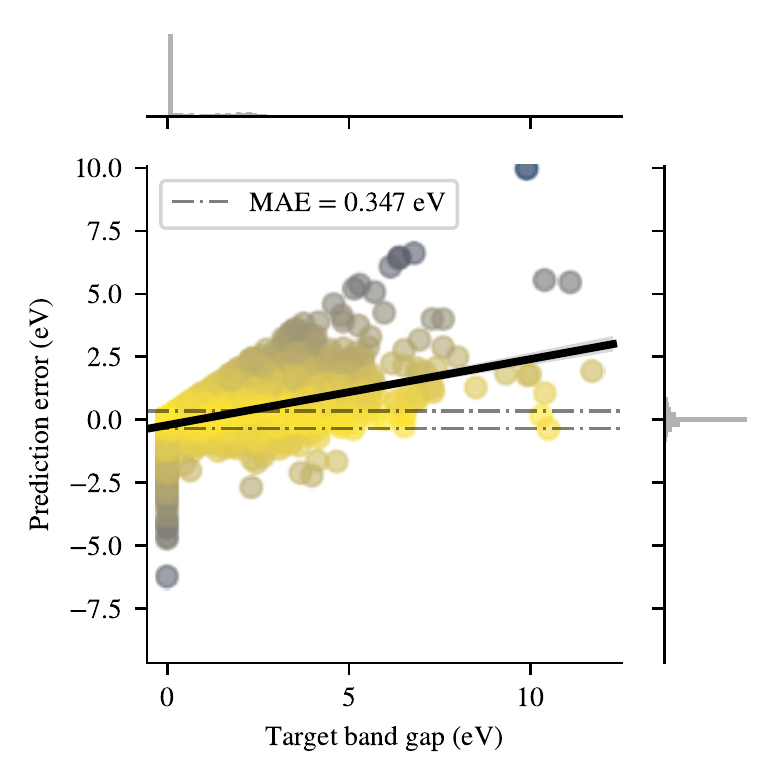}
    \caption{Results for the \texttt{matbench\_expt\_gap} dataset.}
    \label{fig:matbench_expt_gap}
\end{figure}

\begin{figure}[h]
    \centering
    \includegraphics[width=\textwidth]{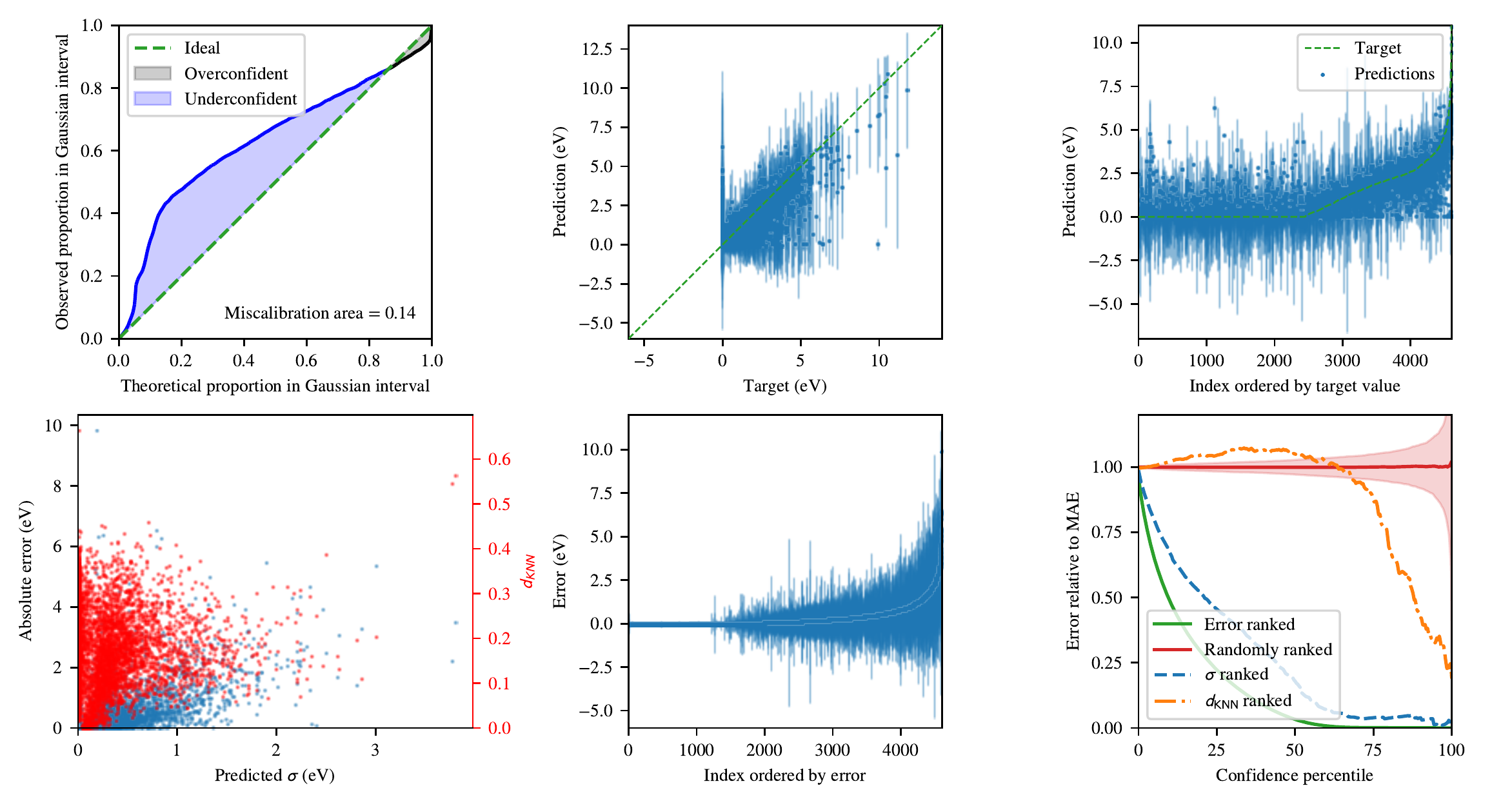}
    \caption{Uncertainty results for the \texttt{matbench\_expt\_gap} dataset.}
    \label{fig:unc_expt_gap}
\end{figure}
\clearpage

\subsection{\texttt{matbench\_jdft2d}}
\begin{figure}[h]
    \centering
    \includegraphics[width=0.495\textwidth]{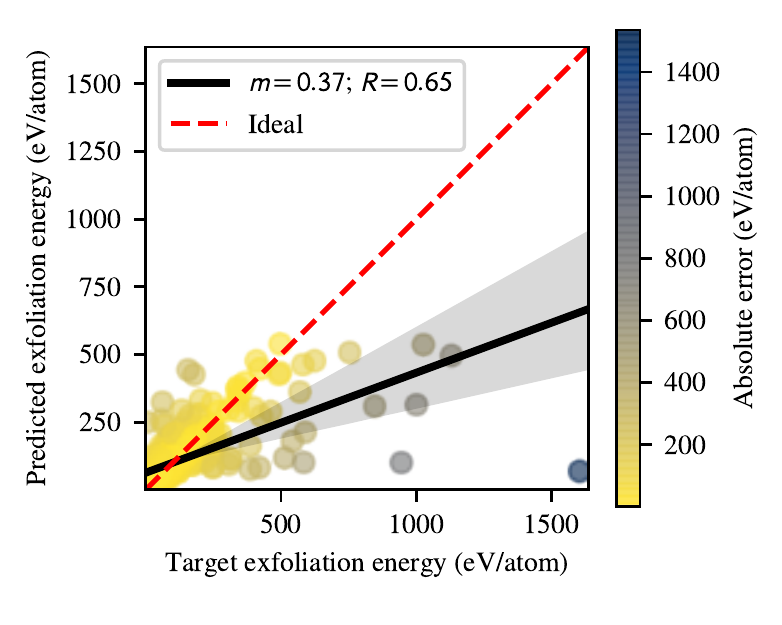}
    \includegraphics[width=0.495\textwidth]{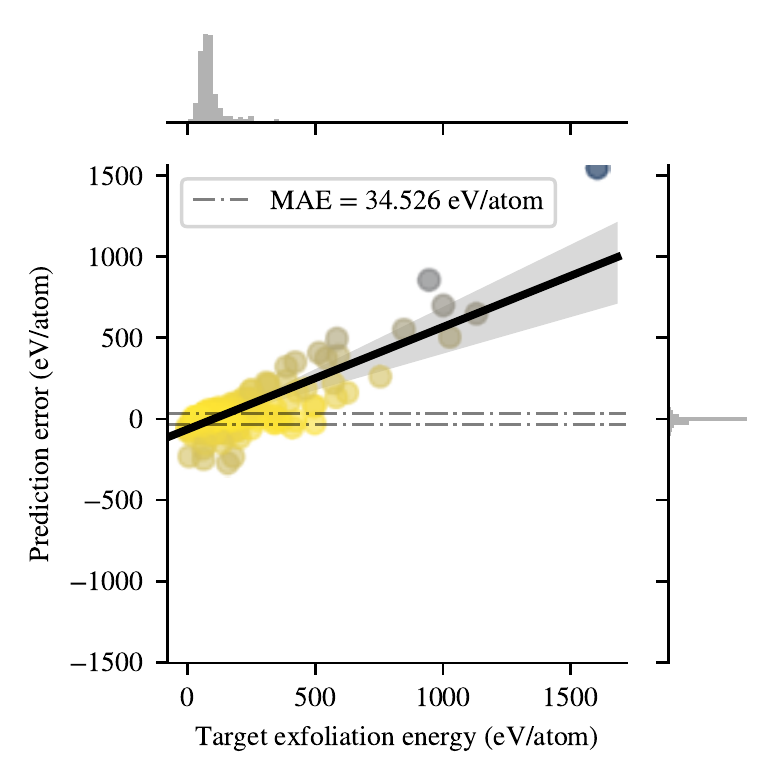}
    \caption{Results for the \texttt{matbench\_jdft2d} dataset.}
    \label{fig:matbench_jdft2d}
\end{figure}

\begin{figure}[h]
    \centering
    \includegraphics[width=\textwidth]{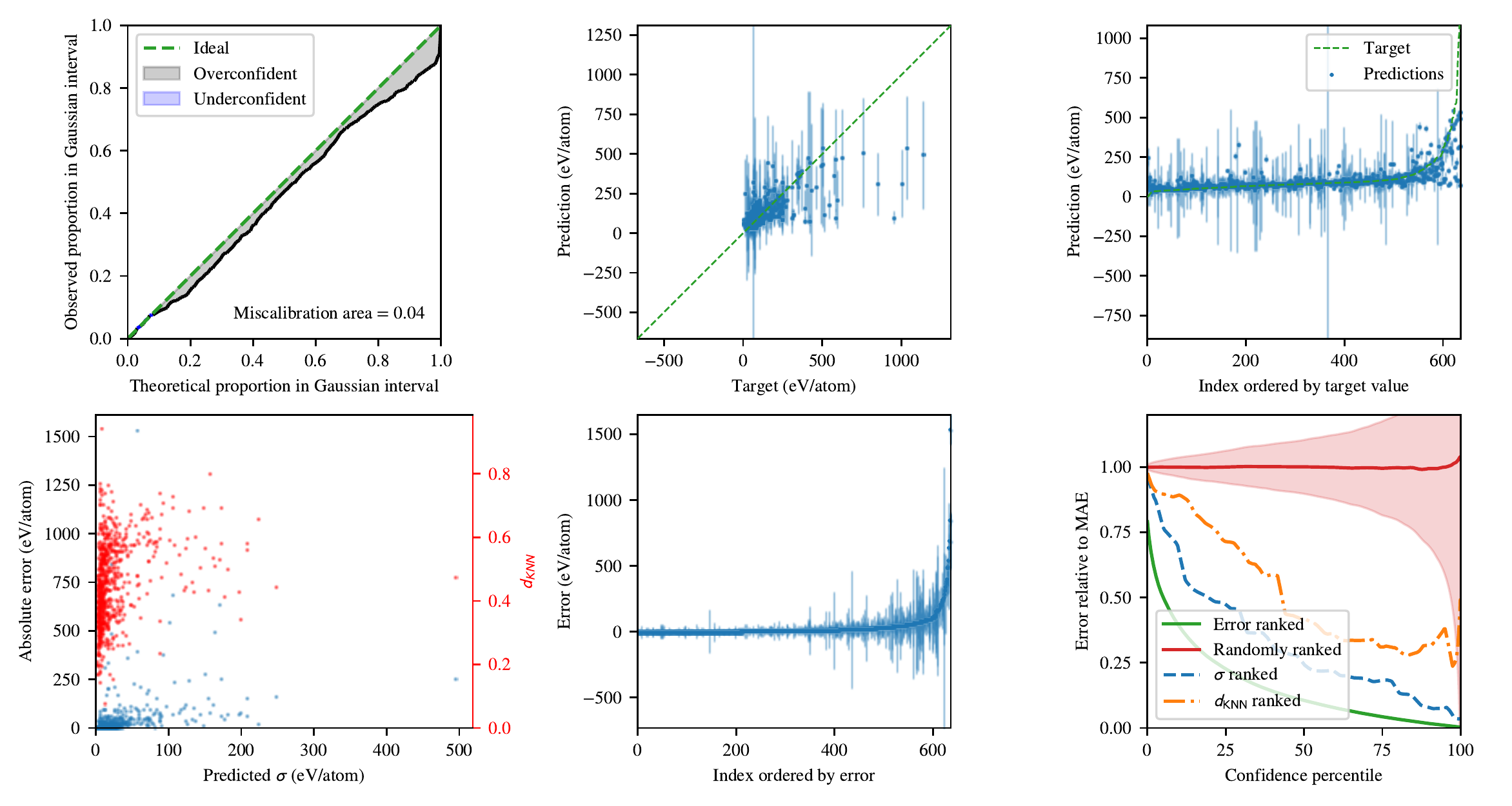}
    \caption{Uncertainty results for the \texttt{matbench\_jdft2d} dataset.}
    \label{fig:unc_jdft2d}
\end{figure}
\clearpage

\subsection{\texttt{matbench\_log\_gvrh}}
\begin{figure}[h]
    \centering
    \includegraphics[width=0.495\textwidth]{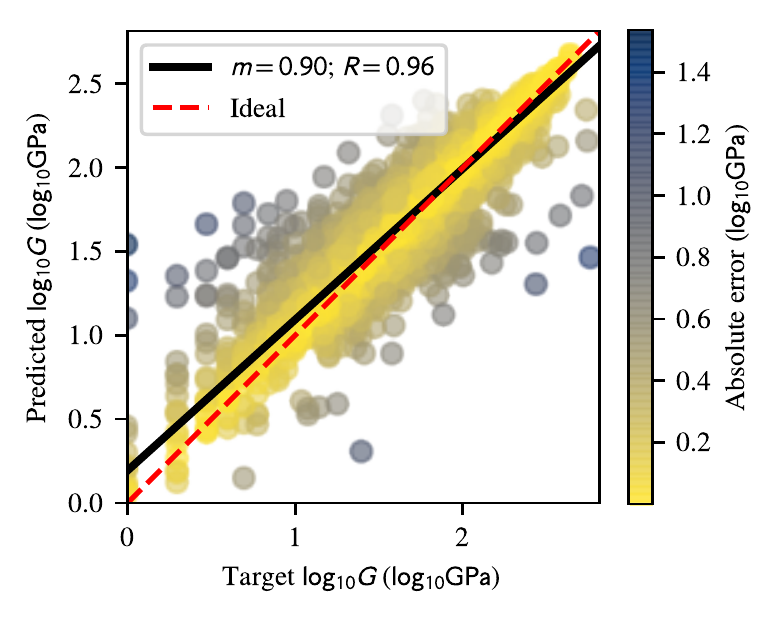}
    \includegraphics[width=0.495\textwidth]{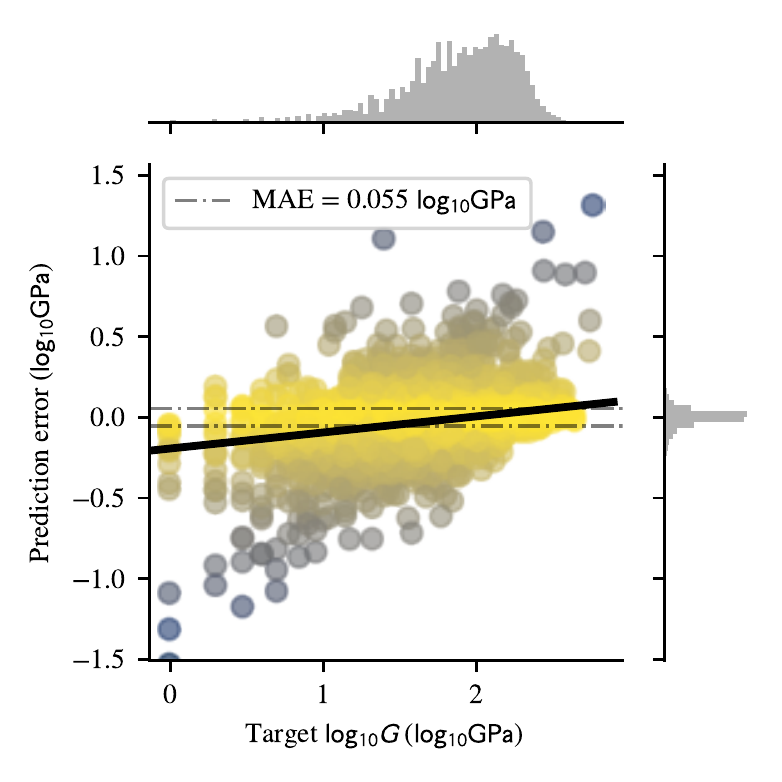}
    \caption{Results for the \texttt{matbench\_log\_gvrh} dataset.}
    \label{fig:matbench_log_gvh}
\end{figure}

\begin{figure}[h]
    \centering
    \includegraphics[width=\textwidth]{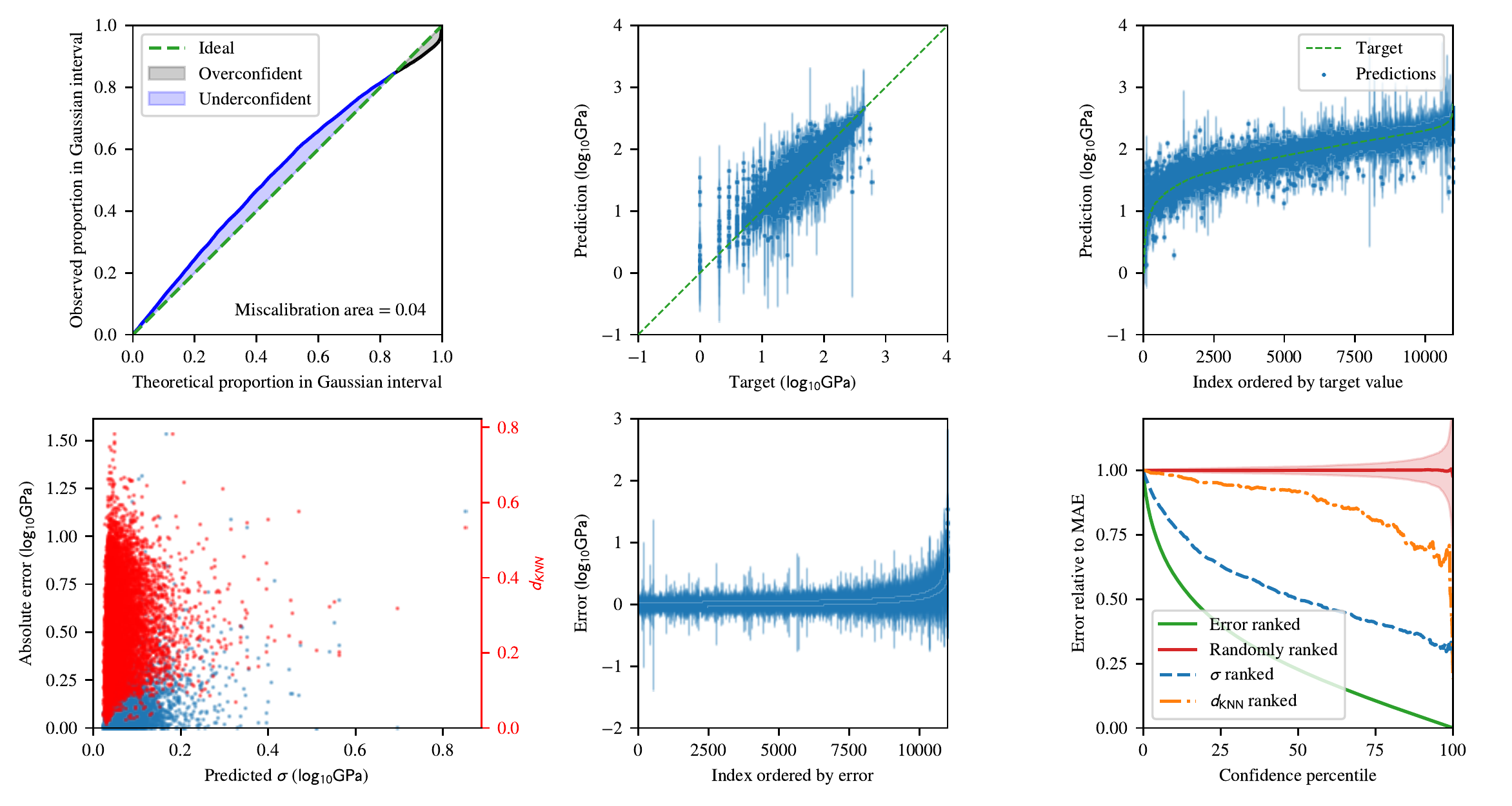}
    \caption{Uncertainty results for the \texttt{matbench\_log\_gvrh} dataset.}
    \label{fig:unc_log_gvh}
\end{figure}
\clearpage
\subsection{\texttt{matbench\_log\_kvrh}}
\begin{figure}[h]
    \centering
    \includegraphics[width=0.495\textwidth]{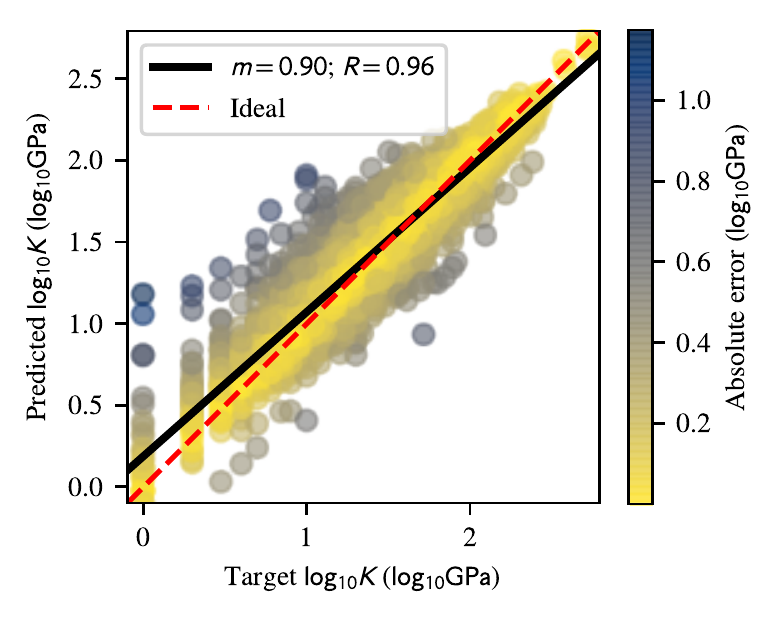}
    \includegraphics[width=0.495\textwidth]{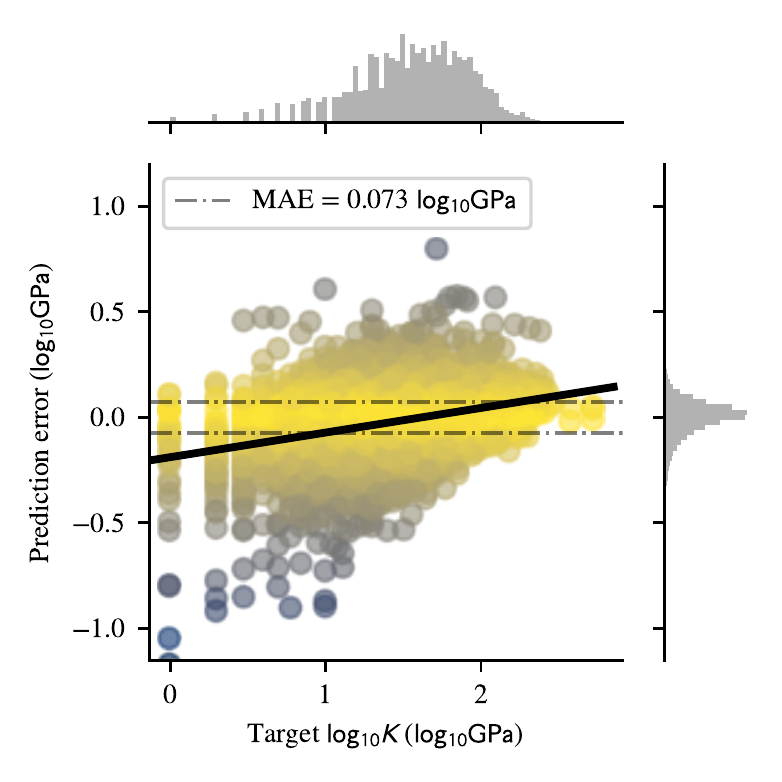}
    \caption{Results for the \texttt{matbench\_log\_kvrh} dataset.}
    \label{fig:matbench_log_kvrh}
\end{figure}

\begin{figure}[h]
    \centering
    \includegraphics[width=\textwidth]{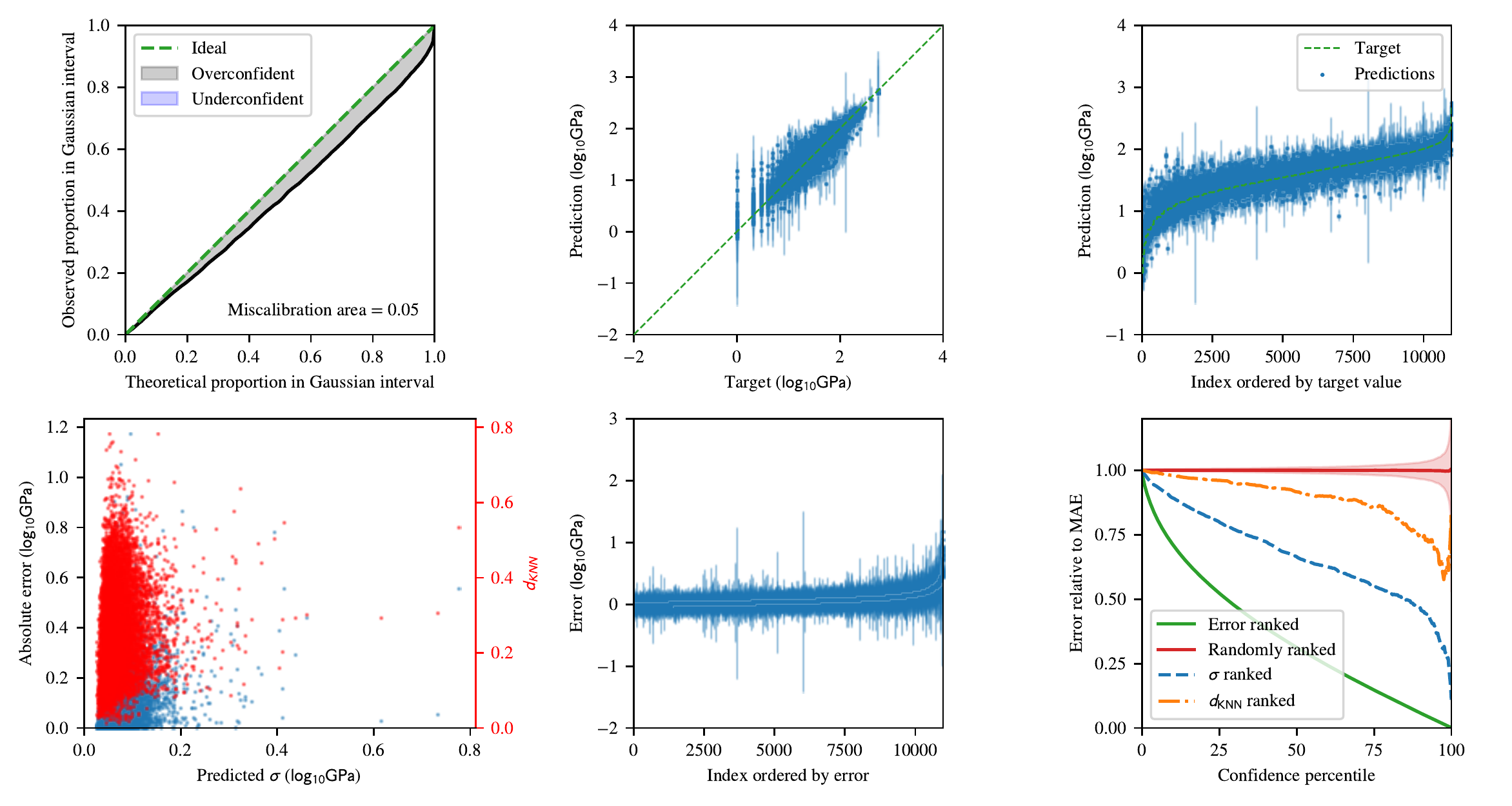}
    \caption{Uncertainty results for the \texttt{matbench\_log\_kvrh} dataset.}
    \label{fig:unc_log_kvrh}
\end{figure}
\clearpage

\subsection{\texttt{matbench\_phonons}}
\begin{figure}[h]
    \centering
    \includegraphics[width=0.495\textwidth]{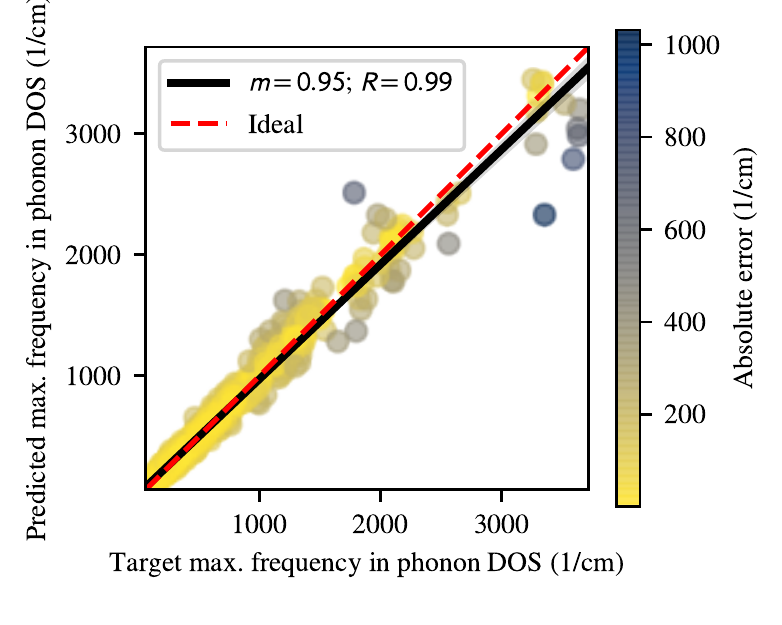}
    \includegraphics[width=0.495\textwidth]{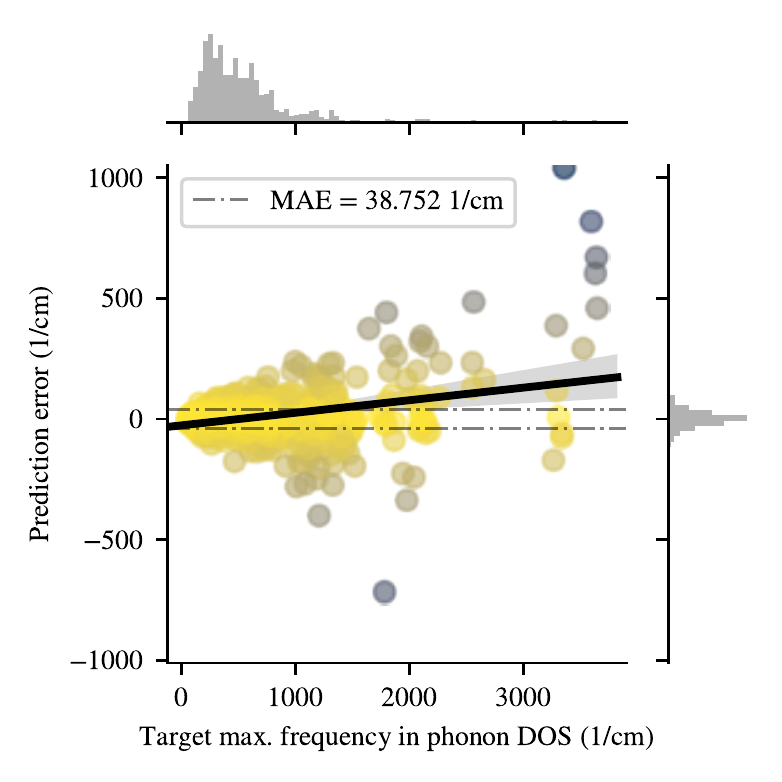}
    \caption{Results for the \texttt{matbench\_phonons} dataset.}
    \label{fig:matbench_phonons}
\end{figure}
\begin{figure}[h]
    \centering
    \includegraphics[width=\textwidth]{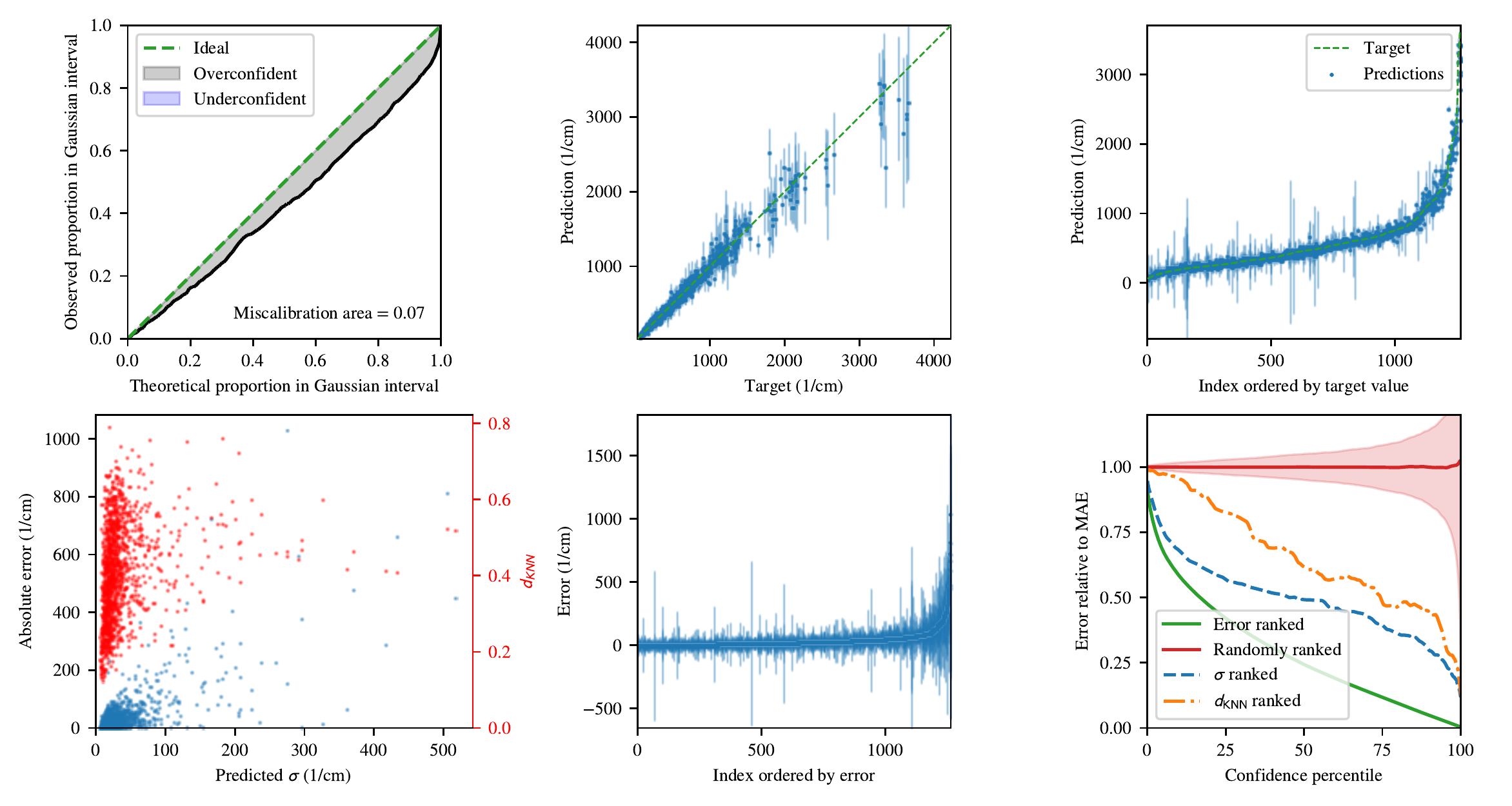}
    \caption{Uncertainty results for the \texttt{matbench\_phonons} dataset.}
    \label{fig:unc_phonons}
\end{figure}

\clearpage

\subsection{\texttt{matbench\_steels}}
\begin{figure}[h]
    \centering
    \includegraphics[width=0.495\textwidth]{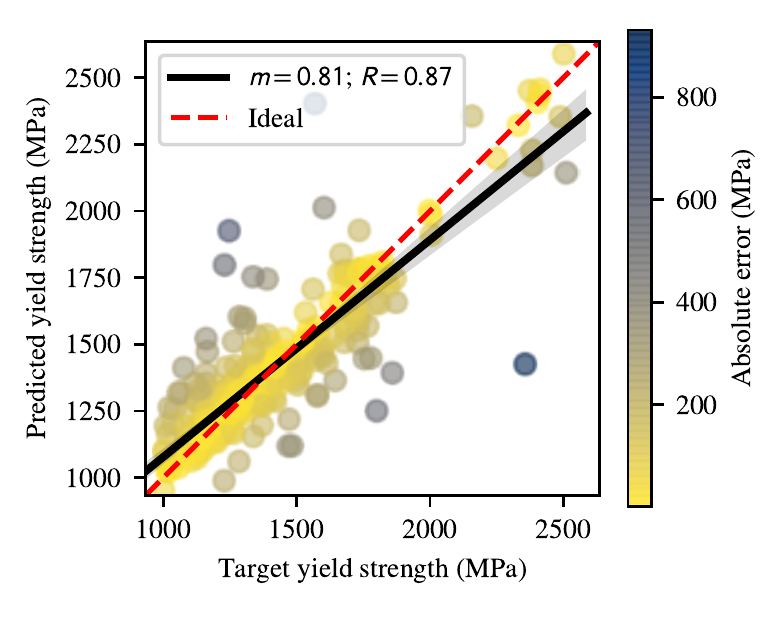}
    \includegraphics[width=0.495\textwidth]{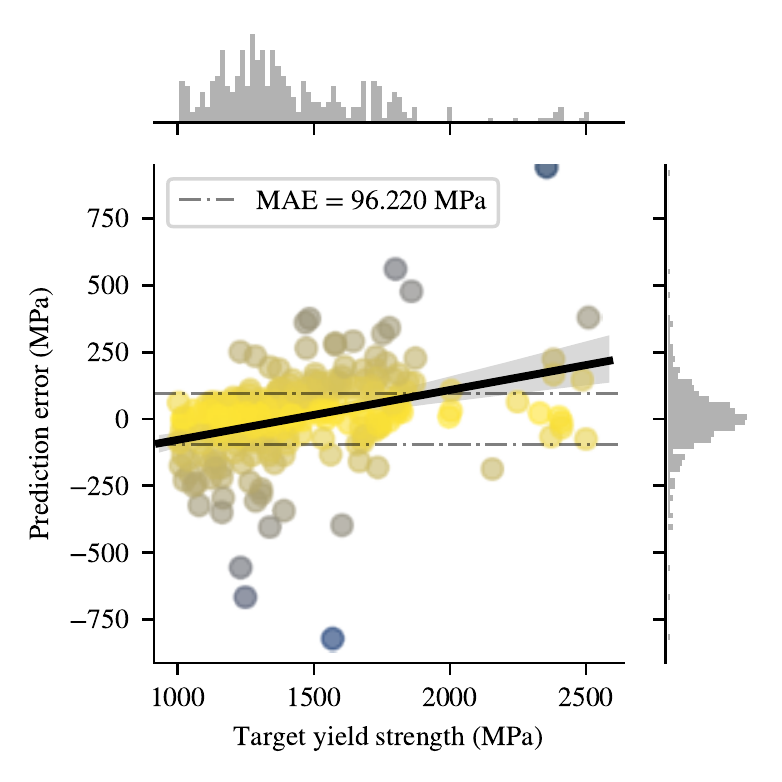}
    \caption{Results for the \texttt{matbench\_steels} dataset.}
    \label{fig:matbench_steels}
\end{figure}
\begin{figure}[h]
    \centering
    \includegraphics[width=\textwidth]{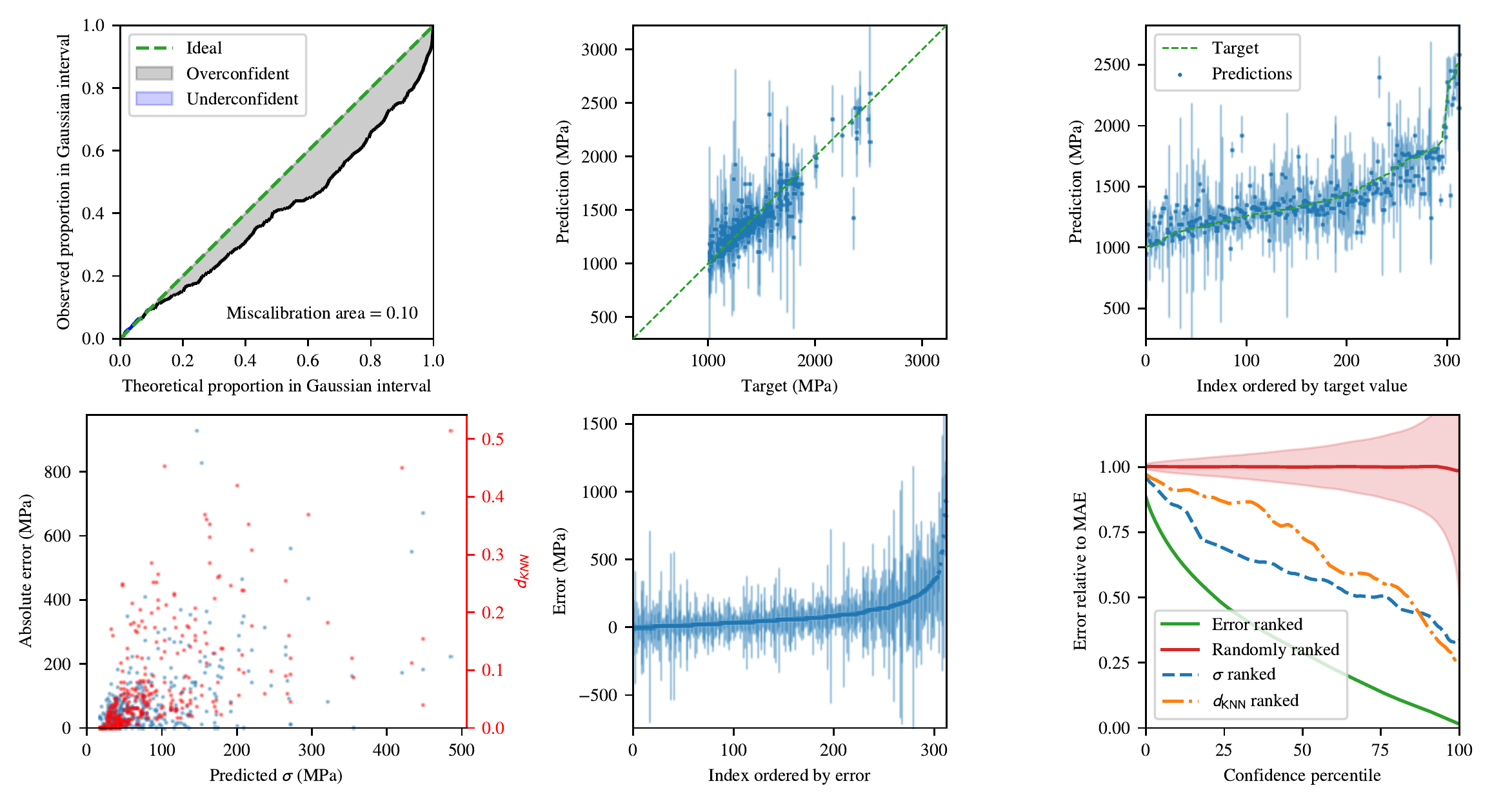}
    \caption{Uncertainty results for the \texttt{matbench\_steels} dataset.}
    \label{fig:unc_steels}
\end{figure}
\clearpage

\subsection{\texttt{matbench\_expt\_is\_metal}}

\begin{figure}[h]
    \centering
    \includegraphics[width=0.95\textwidth]{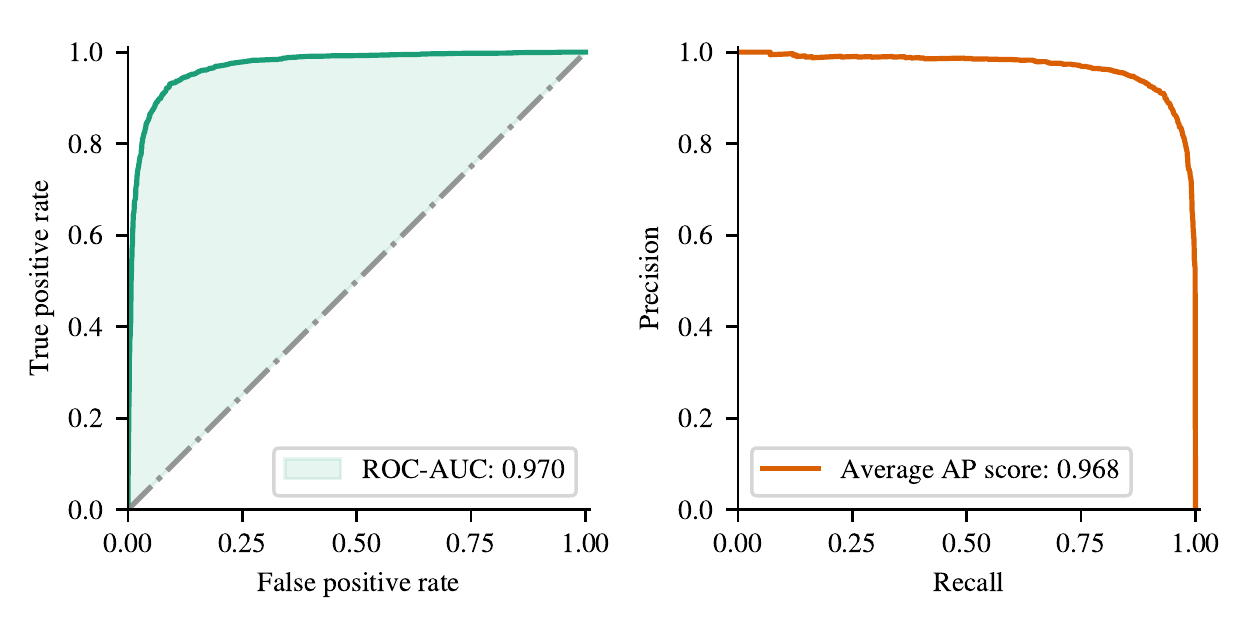}
    \caption{Results for the \texttt{matbench\_expt\_is\_metal} dataset.}
    \label{fig:matbench_expt_is_metal}
\end{figure}

\subsection{\texttt{matbench\_glass}}
\begin{figure}[h]
    \centering
    \includegraphics[width=0.95\textwidth]{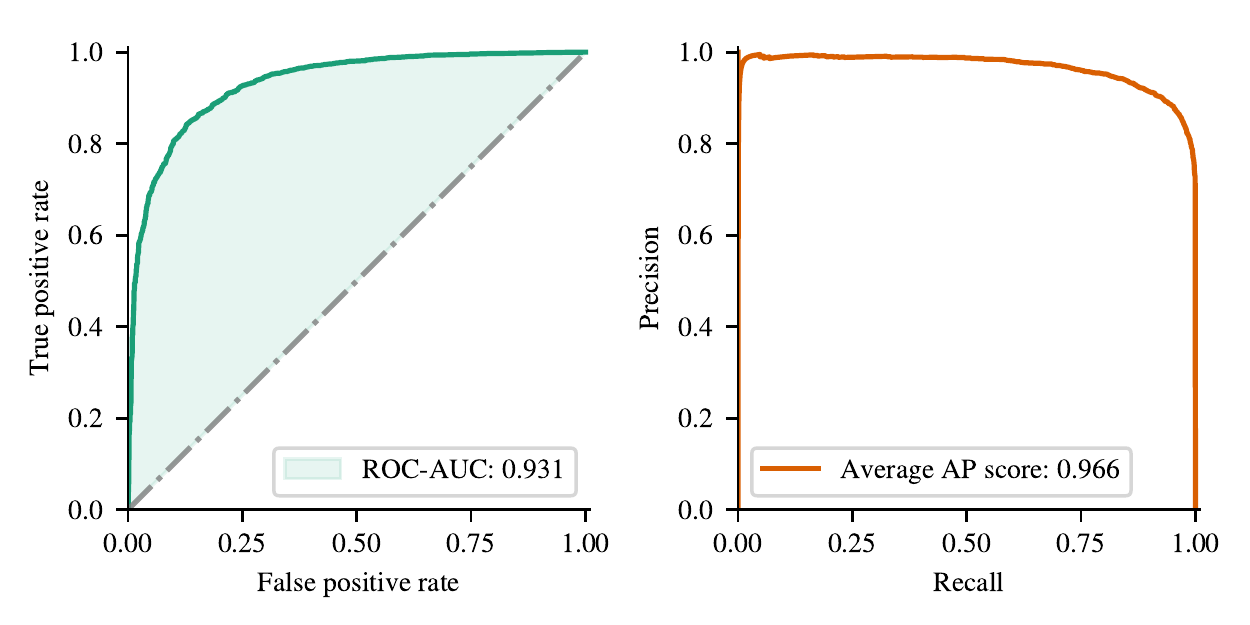}
    \caption{Results for the \texttt{matbench\_glass} dataset.}
    \label{fig:matbench_glass}
\end{figure}

\clearpage

\section{PCA}
\label{appendix:pca}

\subsection{First component (PC$_1$)}
$$PC_1 = \sum w_{1,i} f_{1,i}$$

\begin{table}[h!]
\caption{20 highest contributing features of PC$_1$ with corresponding weights}
\begin{indented}
\item[]\begin{tabular}{@{}cc}
\br
$w_{1,i}$ & Feature $f_{1,i}$\\
\hline
-0.0890 & \texttt{ChemEnvSiteFingerprint|mean SC:12} \\
-0.0890 & \texttt{ChemEnvSiteFingerprint|mean SH:11} \\
-0.0890 & \texttt{ChemEnvSiteFingerprint|std\_dev S:10} \\
-0.0890 & \texttt{ChemEnvSiteFingerprint|mean DD:20} \\
-0.0890 & \texttt{ChemEnvSiteFingerprint|mean H:10} \\
-0.0890 & \texttt{ChemEnvSiteFingerprint|std\_dev CO:11} \\
-0.0890 & \texttt{ChemEnvSiteFingerprint|mean S:12} \\
-0.0890 & \texttt{ChemEnvSiteFingerprint|mean S:10} \\
-0.0890 & \texttt{ChemEnvSiteFingerprint|mean CO:11} \\
-0.0890 & \texttt{ChemEnvSiteFingerprint|std\_dev SH:11} \\
-0.0890 & \texttt{ChemEnvSiteFingerprint|std\_dev S:12} \\
-0.0890 & \texttt{ChemEnvSiteFingerprint|std\_dev H:10} \\
-0.0890 & \texttt{ChemEnvSiteFingerprint|std\_dev DD:20} \\
-0.0890 & \texttt{ChemEnvSiteFingerprint|std\_dev SC:12} \\
-0.0890 & \texttt{ChemEnvSiteFingerprint|mean H:11} \\
-0.0890 & \texttt{ChemEnvSiteFingerprint|mean HD:9} \\
-0.0890 & \texttt{ChemEnvSiteFingerprint|mean SH:13} \\
-0.0890 & \texttt{ChemEnvSiteFingerprint|std\_dev H:11} \\
-0.0890 & \texttt{ChemEnvSiteFingerprint|mean PCPA:11} \\
-0.0890 & \texttt{ChemEnvSiteFingerprint|mean TBSA:10} \\
\br
\end{tabular}
\end{indented}
\end{table}
\clearpage

\subsection{Second component (PC$_2$)}
$$PC_2 = \sum w_{2,i} f_{2,i}$$

\begin{table}[h!]
\caption{20 highest contributing features of PC$_2$ with corresponding weights}
\begin{indented}
\item[]\begin{tabular}{@{}cc}
\br
$w_{2,i}$ & Feature $f_{2,i}$\\
\hline
-0.1070 & \texttt{GaussianSymmFunc|mean G2\_20.0} \\
-0.1053 & \texttt{AGNIFingerPrint|mean AGNI eta=1.23e+00} \\
-0.1049 & \texttt{GeneralizedRDF|mean Gaussian center=1.0 width=1.0} \\
0.1047 & \texttt{ElementProperty|MagpieData mean Row} \\
0.0996 & \texttt{VoronoiFingerprint|mean Voro\_dist\_minimum} \\
-0.0993 & \texttt{GeneralizedRDF|mean Gaussian center=0.0 width=1.0} \\
-0.0973 & \texttt{AGNIFingerPrint|std\_dev AGNI eta=1.23e+00} \\
0.0966 & \texttt{AverageBondLength|mean Average bond length} \\
-0.0965 & \texttt{AGNIFingerPrint|mean AGNI eta=1.88e+00} \\
0.0962 & \texttt{ElementProperty|MagpieData mean Number} \\
-0.0949 & \texttt{GeneralizedRDF|std\_dev Gaussian center=0.0 width=1.0} \\
-0.0948 & \texttt{GaussianSymmFunc|std\_dev G2\_20.0} \\
0.0948 & \texttt{ElementProperty|MagpieData mean CovalentRadius} \\
0.0945 & \texttt{ElementProperty|MagpieData mean AtomicWeight} \\
-0.0877 & \texttt{GaussianSymmFunc|mean G4\_0.005\_4.0\_-1.0} \\
-0.0876 & \texttt{AGNIFingerPrint|std\_dev AGNI dir=y eta=1.23e+00} \\
-0.0864 & \texttt{AGNIFingerPrint|std\_dev AGNI dir=x eta=1.23e+00} \\
-0.0859 & \texttt{GaussianSymmFunc|mean G2\_4.0} \\
-0.0848 & \texttt{AGNIFingerPrint|std\_dev AGNI dir=y eta=1.88e+00} \\
-0.0845 & \texttt{GeneralizedRDF|std\_dev Gaussian center=1.0 width=1.0} \\
\br
\end{tabular}
\end{indented}
\end{table}
\clearpage
\subsection{Third component (PC$_3$)}
$$PC_3 = \sum w_{3,i} f_{3,i}$$

\begin{table}[h!]
\caption{20 highest contributing features of PC$_3$ with corresponding weights}
\begin{indented}
\item[]\begin{tabular}{@{}cc}
\br
$w_{3,i}$ & Feature $f_{3,i}$\\
\hline
-0.1194 & \texttt{IonProperty|avg ionic char} \\
-0.1174 & \texttt{ElementProperty|MagpieData avg\_dev Electronegativity} \\
-0.1159 & \texttt{LocalPropertyDifference|mean local difference in Electronegativity} \\
-0.1023 & \texttt{IonProperty|max ionic char} \\
0.1020 & \texttt{ElementProperty|MagpieData mean MendeleevNumber} \\
-0.1005 & \texttt{ElementProperty|MagpieData avg\_dev CovalentRadius} \\
-0.0995 & \texttt{DensityFeatures|packing fraction} \\
-0.0990 & \texttt{ElementProperty|MagpieData range Electronegativity} \\
-0.0959 & \texttt{ElectronegativityDiff|mean EN difference} \\
-0.0957 & \texttt{ElementProperty|MagpieData avg\_dev MendeleevNumber} \\
-0.0955 & \texttt{ElementProperty|MagpieData avg\_dev SpaceGroupNumber} \\
-0.0950 & \texttt{ElementProperty|MagpieData avg\_dev Column} \\
0.0950 & \texttt{ElementProperty|MagpieData minimum Electronegativity} \\
0.0939 & \texttt{ElementProperty|MagpieData minimum Column} \\
0.0936 & \texttt{ElementProperty|MagpieData mean NpValence} \\
0.0936 & \texttt{ValenceOrbital|avg p valence electrons} \\
0.0935 & \texttt{ElementProperty|MagpieData minimum MendeleevNumber} \\
-0.0922 & \texttt{ElectronegativityDiff|maximum EN difference} \\
0.0898 & \texttt{ElementProperty|MagpieData mean Column} \\
-0.0896 & \texttt{ElementProperty|MagpieData range Column} \\
\br
\end{tabular}
\end{indented}
\end{table}

\end{document}